# Theory of Electron-Phonon-Dislon Interacting System – Toward a Quantized Theory of Dislocations

Mingda Li[1,2*], Yoichiro Tsurimaki[1], Qingping Meng[3], Nina Andrejevic[4], Yimei Zhu[3], Gerald D. Mahan[5] and Gang Chen[1*]

[1]*Department of Mechanical Engineering, MIT, Cambridge, MA 02139, USA*

[2]*Department of Nuclear Science and Engineering, MIT, Cambridge, MA 02139, USA*

[3]*Condensed Matter Physics and Material Sciences Department, Brookhaven National Laboratory, Upton, NY 11973, USA*

[4]*Department of Materials Science and Engineering, MIT, Cambridge, MA 02139, USA*

[5]*Department of Physics, The Pennsylvania State University, University Park, PA 16802, USA*

We provide a comprehensive theoretical framework to study how crystal dislocations influence the functional properties of materials, based on the idea of a quantized dislocation, namely a "dislon". In contrast to previous work on dislons which focused on exotic phenomenology, here we focus on their theoretical structure and computational power. We first provide a pedagogical introduction that explains the necessity and benefits of taking the dislon approach and why the dislon Hamiltonian takes its current form. Then, we study the electron-dislocation and phonon-dislocation scattering problems using the dislon formalism. Both the effective electron and phonon theories are derived, from which the role of dislocations on electronic and phononic transport properties is computed. Compared with traditional dislocation scattering studies, which are intrinsically single-particle, low-order perturbation and classical quenched defect in nature, the dislon theory not only allows easy incorporation of quantum many-body effects such as electron correlation, electron-phonon interaction, and higher-order scattering events, but also allows proper consideration of the dislocation's long-range strain field and dynamic aspects on equal footing for arbitrary types of straight-line dislocations. This means that instead of developing individual models for specific dislocation scattering problems, the dislon theory allows for the calculation of electronic structure and electrical transport, thermal transport, optical and superconducting properties, etc., under one unified theory. Furthermore, the dislon theory has another advantage over empirical models in that it requires no fitting parameters. The dislon theory could serve as a major computational tool to understand the role of dislocations on multiple materials' functional properties at an unprecedented level of clarity, and may have wide applications in dislocated energy materials.

## I. INTRODUCTION

Dislocations are a common type of line defect in crystalline solids [1]. Since the first theoretical predictions of dislocations in 1934 [2-4] and experimental observations in 1956 [5-7], most dislocation studies have focused on their impacts on materials' mechanical properties, where a pure classical description of dislocations has been proven highly successful [1,8,9]. In addition to their influence on mechanical properties, dislocations are also known to affect a number of materials' functional properties such as electronic structure and electrical transport [10-25], thermal transport [26-36], optical properties [37-51], magnetic properties [52-63], and even superconductivity [64-76]. However, in contrast to the well-studied mechanical properties based on the classical dislocation framework, the theoretical studies of functional properties have long been restricted to numerous but scattered empirical dislocation models. In each specific case, a dislocation is modeled as a certain object and often accompanied by empirical parameters in order to fit experimental data. On the other hand, given the rapid development of novel functional materials and exotic condensed matter phases, alongside the advancement of nanoscience and device miniaturization, the interplay between dislocations and the various novel functional properties is simply beyond the grasp of those case-by-case, oversimplified empirical models. For instance, conventional dislocation theory becomes completely helpless if we ask simple questions like "*should dislocations increase or decrease the critical temperature of a dislocated superconductor*", "*how do dislocations change magnetic order or optical spectra*" or "*can dislocations drive a band insulator into a topological insulator*", etc. In this sense, it is unimaginable to develop a separate model for each scenario.

Recently, some of us started to develop a theory based on the quantization of a dislocation and introduced the "dislon" as the basic quanta of a quantized dislocation for arbitrary types of dislocation lines, including both edge and screw dislocations. Starting from a one-dimensional quantization approach [25,36,77], we treated electron [25] and phonon [36] scattering with an individual dislocation, and single electron-interacting dislocation pair scattering [77]. Later, we generalized the one-dimensional approach to a full three-dimensional quantization [76], but focused on the electron-dislocation superconductivity.

In this study, we generalize the dislon theory in 3D to include electron and phonon interactions with multiple dislocations. To demonstrate the possible utility of the dislon theory, we derive expressions for transport properties such as the electron relaxation time of electron-dislon scattering, the electrical conductivity caused by electron-dislocation scattering, and the thermal conductivity caused by phonon-dislocation interaction. The structure of this paper is as follows. In Sec. II, we provide a heuristic argument for dislons. Sections III-V are the full quantization procedure and free dislon theory. Sections VI-VIII are the Hamiltonian theories for electron-dislon and phonon dislon interactions, from which the effective electron and phonon theories after eliminating the dislon degrees of freedom are derived in Sections IX-XII. A significant portion of Sections VI and XII has been reported in [76]and is rewritten here in greater detail to keep the study self-contained. In Sections XI-XV, the electronic and phononic transport properties are derived, followed by discussions, future perspective, and conclusions in Sections XVI-XVII.

## II. HEURISTIC ARGUMENT

First, we would like to briefly introduce the motivation behind the theory: We plan to develop a unified, microscopic, and quantitative theory of dislocations which can be applied to calculate materials' functional properties. However, one question remains: why does a quantized dislocation approach have to be adopted to fulfill this goal? To see this, we should be aware that all the functional properties – no matter electrical, magnetic, optical, thermoelectric or superconducting properties – are intrinsically quantum properties, and can be well described by quantum many-body theories microscopically. Therefore, a quantum description of dislocations can easily be integrated into the modern many-body formalism, thereby taking full advantage of all theoretical field techniques. This motivates development of the "dislon" – the quantized dislocation theory.

Despite strong motivation for introducing the idea of a quantized dislocation, what is the justification of the approach? In other words, why is a classical, quenched dislocation approach insufficient? To see this, we need to bear in mind that a dislocation is actually NOT a simple quenched defect. Taking phonons as an example, when a phonon scatters with a dislocation, it is widely accepted that the dominant interaction mechanism is scattering from a vibrating dislocation, called "fluttering" [29,31,32,78]. In fact, both theories [36,79] and simulation [80] have shown the existence of resonance of the dislocation-phonon interaction, indicating a similar dynamic energy scale between a phonon and a dislocation which cannot be captured by any static, quenched disorder model. The electron-phonon scattering mechanism [81], together with the dislocation-phonon resonance, indicates the possible role that a dynamic dislocation may also play in electron-dislocation scattering problems.

On the other hand, a quantized dislocation, aka "dislon", can incorporate both quenched dislocation effects and all the dynamic features on equal footing. Just as phonons are quantized fluctuations around a perfect periodic crystal structure, dislons are also quantized fluctuations around classical quenched dislocations. More rigorously, defining $\mathbf{u}$ as the lattice displacement vector (deviation from the equilibrium position), phonons are then cast as quantized propagating vibrational modes by rewriting $\mathbf{u}$ in terms of phonon creation and annihilation operators, while dislons are cast as quantized localized modes by writing $\mathbf{u}$ in terms of dislon creation and annihilation operators, but further satisfying a dislocation's rigorous definition $\oint_D d\mathbf{u} = -\mathbf{b}$, in which $\mathbf{b}$ is the Burgers vector and $D$ is an arbitrary loop enclosing the dislocation line. By taking this rigorous definition as the classical limit, all classical effects of dislocations are automatically incorporated, but most importantly, a whole new territory to study the quantum effects of dislocations on materials also unfolds (imagine the huge difference between a classical elastic wave and a quantized phonon).

Prior to the formal derivation, we briefly argue the existence of a certain form of excitation beyond phonon excitation in a dislocated crystal from the perspective of eigenmode conservation. Assuming a perfect periodic solid containing $N$ atoms, which has $3N$ eigenmodes labeled by $3N$ good quantum numbers called crystal momentum $\mathbf{k}$. Assume the Debye frequency in this solid is $\omega_D$; then, for a solid with 1 atom per unit cell in which only acoustic phonons exist, the Debye model tells us that [82]

$$3N = \sum_{\mathbf{k}} 1 = \int_0^{\omega_D} D_{ph}(\omega) d\omega \tag{II.1}$$

where $D_{ph}(\omega)$ is the phonon density of states, which can be written as

$$D_{ph}(\omega) = \sum_{\mathbf{k}} \delta(\omega - \varepsilon_{\mathbf{k}}) \tag{II.2}$$

For a simple acoustic phonon, $\omega = v_{\mathrm{ph}} k$, where $v_{\mathrm{ph}}$ is speed of the acoustic phonon, and it can be shown directly that $D_{ph}(\omega) = L^3\omega^2 / 2\pi^2 v_{\mathrm{ph}}^3 = 9N\omega^2/\omega_D^3$ is valid, where $L$ is the sample size.

Now with dislocations present, crystal momenta $\mathbf{k}$ are no longer rigorously good quantum numbers, indicating a reduction in the phonon density of states $D_{ph}(\omega)$ which represents the portion of extended, propagating eigenmodes which in general preserve translational symmetry. However, the total number of modes should still be conserved, giving

$$3N = \int_0^{\omega_D} D_{tot}(\omega) d\omega = \int_0^{\omega_D} D_{ph}(\omega) d\omega + \int_0^{\omega_D'} D'(\omega) d\omega \tag{II.3}$$

where an excess density of states $D'(\omega)$ emerges to compensate the reduction of $D_{ph}(\omega)$, and can be intuitively treated as some localized modes. Despite the lack of a simple way to directly separate the $D_{ph}(\omega)$ from $D'(\omega)$, in the case of a dislocated crystal, the appearance of $D'(\omega)$ can be regarded as a result of translational symmetry breaking and the formation of local modes, which gives a qualitative, heuristic rationale for the possible existence of a "dislon" excitation.

With this intuitive understanding in hand, we now formally introduce the dislon theory.

## III. THE THEORETICAL FOUNDATION

First, we define the 3D Cartesian coordinate $\mathbf{R} \equiv (x, y, z)$. Assuming a dislocation line is extending along the $z$-direction, we also define the 2D coordinate $\mathbf{r} \equiv (x, y)$ and assume the dislocation core is located at $\mathbf{r}_0 \equiv (x_0, y_0)$. $\mathbf{r}_0$ is a 2D vector since a dislocation is a line defect. For a single dislocation line, we define $\mathbf{u}(\mathbf{R})$ as the lattice displacement field at the spatial position $\mathbf{R}$ caused by the dislocation with dislocation core location $\mathbf{r}_0$. We further define the 3D momentum coordinate $\mathbf{k} = (k_x, k_y, k_z)$, 2D momentum $\mathbf{s} = (k_x, k_y)$, and crystal size as $L$. We separate $x$ and $y$ from the $z$ direction by defining $\kappa \equiv k_z$ to emphasize the special direction along the dislocation line direction (hence $\mathbf{k} \equiv (\mathbf{s}, \kappa)$). The displacement field $\mathbf{u}(\mathbf{R})$ can be written as a generic form of mode expansion by a generic mode $\mathbf{U}_\mathbf{k}$ (the $1/L^2$ prefactor is for later convenience)

$$\mathbf{u}(\mathbf{R}) = \frac{1}{L^2} \sum_{\mathbf{k}} e^{i\mathbf{k}\cdot\mathbf{R}} e^{-i\mathbf{s}\cdot\mathbf{r}_0} \mathbf{U}_\mathbf{k} \qquad \text{(III.1)}$$

Under the static, or equivalently the long-wavelength, limit, Eq. (III.1) should reduce to the displacement field of a classical quenched dislocation.

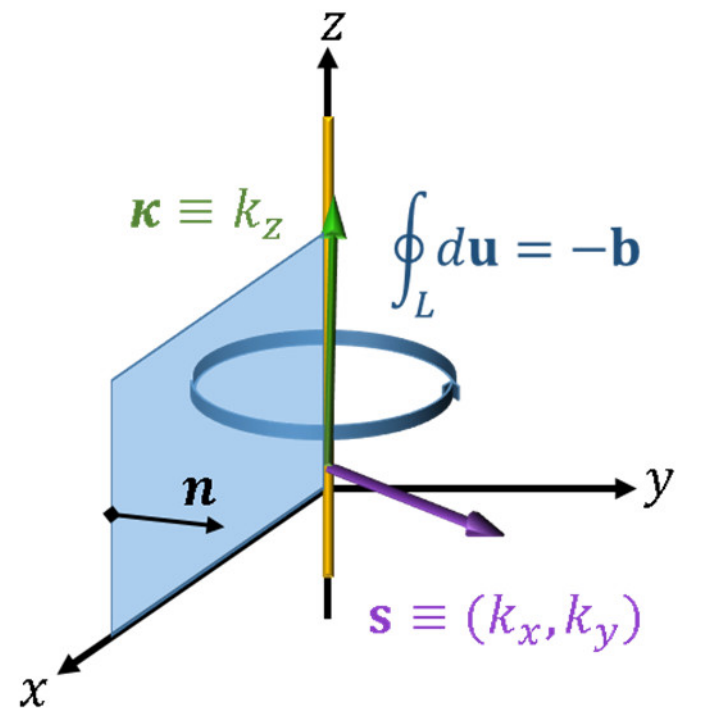

**Fig. 1.** The coordinate system used for this study. The golden line along $z$-direction is the dislocation line direction, and the slip plane is the $xz$ plane. Since a dislocation is a line defect, the momentum component along the dislocation line is highlighted with a different label $\kappa$, which shows later convenience.

To see how a quenched dislocation satisfying the definition $\oint_D d\mathbf{u} = -\mathbf{b}$ can be recovered from Eq. (III.1), we compare the displacement of a dislocation with that of a phonon. The phonon displacement field $\mathbf{u}_{ph}(\mathbf{R})$ can be written as a mode expansion using a plane wave basis as [83]

$$\mathbf{u}_{ph}(\mathbf{R}) = \frac{1}{L^{3/2}} \sum_{\mathbf{k}} e^{i\mathbf{k}\cdot\mathbf{R}} \mathbf{u}_{ph,\mathbf{k}} \qquad \text{(III.2)}$$

Under the static limit, there is no displacement for an acoustic phonon, i.e. $\mathbf{u}_{ph}(\mathbf{R})\big|_{\text{stat}} = 0$ (perfect periodic lattice), where "$\big|_{\text{stat}}$" means taking a static, long-wavelength limit. In the momentum space representation, this static condition can be denoted as $\mathbf{u}_{ph,\mathbf{k}}\big|_{\text{stat}} = \frac{1}{L^{3/2}} \int d\mathbf{R} e^{-i\mathbf{k}\cdot\mathbf{R}} \mathbf{u}_{ph}(\mathbf{R})\big|_{\text{stat}} = 0$. Since the right-hand side is 0 and is independent of $\mathbf{k}$, we can consider the static limit as a long-wavelength $\mathbf{k} \to 0$ limit. This gives a mathematical expression for the static limit of an acoustic phonon: $\lim_{\mathbf{k}\to 0} \mathbf{u}_{ph,\mathbf{k}} = 0$, for $\forall \mathbf{k}$, which can be regarded as a boundary condition. Since this boundary condition is trivial, it is usually overlooked in phonon studies.

When a dislocation is present, the displacement field still exists under the static limit, since even a classical dislocation without any fluctuations can still create lattice displacement. Thus, this quenched classical static dislocation $\mathbf{u}(\mathbf{r})\big|_{\text{stat}}$ can also be mode-expanded using a 2D wavevector $\mathbf{s}$ as

$$\mathbf{u}(\mathbf{r})\big|_{\text{stat}} = \frac{1}{L^2} \sum_{\mathbf{s}} e^{i\mathbf{s}\cdot\mathbf{r}} e^{-i\mathbf{s}\cdot\mathbf{r}_0} \mathbf{F}(\mathbf{s}) \qquad \text{(III.3)}$$

where $\mathbf{F}(\mathbf{s})$ can be interpreted as a non–plane-wave expansion coefficient to be discussed shortly. In this situation, the constraint of a dislocation's expansion mode $\mathbf{U}_\mathbf{k}$ under the static limit in Eq. (III.1) is clear. As in the phonon case, $\mathbf{U}_\mathbf{k}$ can be defined in a similar way using the inverse Fourier transform as a constraint, whereby

$$\begin{aligned} \mathbf{U}_\mathbf{k}\big|_{\text{stat}} &= \frac{1}{L} \int d\mathbf{R} e^{-i\mathbf{k}\cdot\mathbf{R}} e^{+i\mathbf{s}\cdot\mathbf{r}_0} \mathbf{u}(\mathbf{r})\big|_{\text{stat}} \\ &= \frac{\mathbf{F}(\mathbf{s})}{L} \int dz e^{-i\kappa z} = \mathbf{F}(\mathbf{s}) \delta_{\kappa 0} \end{aligned} \qquad \text{(III.4)}$$

where the right-hand side is now independent of the $z$-component momentum $\kappa$ but still a function of $\mathbf{s}$ (keep in mind $\mathbf{k} \equiv (\mathbf{s}, \kappa)$). This shows that in the displacement field with a dislocation, the static limit is not $\mathbf{k} \to 0$, but involves only the $z$-component of momentum, i.e the $\kappa \to 0$ limit. Thus, the constraint which allows the reduction to a classical quenched dislocation can be written as

$$\lim_{\kappa\to 0} \mathbf{U}_\mathbf{k} \equiv \lim_{\kappa\to 0} \mathbf{U}_{\mathbf{s},\kappa} = \mathbf{F}(\mathbf{s}), \text{ for } \forall \mathbf{s} \qquad \text{(III.5)}$$

At this moment, we should bear in mind that $\mathbf{F}(\mathbf{s})$ is still a function of $\mathbf{s}$ and it should satisfy the dislocation's definition $\oint_D d\mathbf{u}_{\text{stat}}(\mathbf{r}) = -\mathbf{b}$ . A valid form of $\mathbf{F}(\mathbf{s})$ satisfying the definition is provided later in this section.

Up to this point, the entire derivation is valid for a generic *vector* displacement $\mathbf{U}_\mathbf{k}$. To further simplify, we define a generic *scalar* displacement component $u_\mathbf{k}$ as

$$\mathbf{U}_\mathbf{k} \equiv u_\mathbf{k}\mathbf{F}(\mathbf{k}) \qquad \text{(III.6)}$$

where $\mathbf{F}(\mathbf{k})$ is an undetermined 3D function generalized from $\mathbf{F}(\mathbf{s})$ by satisfying $\lim_{\kappa\to 0}\mathbf{F}(\mathbf{k}) = \mathbf{F}(\mathbf{s})$ , aka $\mathbf{F}(\mathbf{k}) \equiv \mathbf{F}(\mathbf{s},\kappa)$ and $\mathbf{F}(\mathbf{s}) \equiv \mathbf{F}(\mathbf{s},0)$, then we have

$$\begin{aligned} &\lim_{\kappa\to 0}\mathbf{U}_\mathbf{k} \equiv \lim_{\kappa\to 0} u_\mathbf{k}\mathbf{F}(\mathbf{k}) = \mathbf{F}(\mathbf{s})\lim_{\kappa\to 0} u_\mathbf{k} = \mathbf{F}(\mathbf{s}) \\ &\Rightarrow \lim_{\kappa\to 0} u_\mathbf{k} = 1 \end{aligned} \qquad \text{(III.7)}$$

The procedure to introduce the scalar displacement $u_\mathbf{k}$ can also be understood from a comparison with phonons. For phonons, there is no need to separately quantize the three components of the displacement vector $\mathbf{u}_{ph,\mathbf{k}}$ , since the phonon displacement field satisfies $\mathbf{u}_{ph,\mathbf{k}} \propto \boldsymbol{\varepsilon}_\mathbf{k}$ , where $\boldsymbol{\varepsilon}_\mathbf{k}$ is the polarization vector obtained by solving a purely classical lattice wave equation. Here, the fact that $\mathbf{U}_\mathbf{k} \propto \mathbf{F}(\mathbf{k})$ indicates that the three vector components of a dislocation are also linked through the function $\mathbf{F}(\mathbf{k})$ ; the $\kappa$–dependence of displacement indicates the existence of wave propagation along the dislocation line direction, from which the dynamic vibration can be taken into account. Therefore, to quantize a dislocation while considering the dynamic aspects, there are two generic approaches:

1) Quantize the vector Fourier component of lattice displacement $\mathbf{U}_\mathbf{k}$ directly, which is an approach of vector-field quantization. A quantization of a vector field has an analog with *photon* quantization, which is a gauge theory [84], and the quantization procedure may be related to the classical affine gauge field of dislocations [85].

2) Quantize the scalar Fourier component of lattice displacement $u_\mathbf{k}$ , defined through $\mathbf{U}_\mathbf{k} \equiv u_\mathbf{k}\mathbf{F}(\mathbf{k})$ , which is much simpler. This is also similar to the *phonon* quantization, where the quantum fluctuation (dynamic effect) along each direction is assumed to be proportional to $\boldsymbol{\varepsilon}_\mathbf{k}$ , the classical eigenvectors of the dynamic matrix. If we see approach 1) as a full gauge theory like quantum electrodynamics (QED), then the approach 2) is like a scalar QED [84], with much simpler structure but still retaining most of the quantum properties.

Therefore, as long as we admit the way in which a phonon is quantized, we could consider Eq. (III.7) a valid and reasonable assumption. The lattice displacement field caused by a generic dislocation $\mathbf{u}(\mathbf{R})$ can be written as

$$\mathbf{u}(\mathbf{R}) = \frac{1}{L^2}\sum_\mathbf{k} e^{i\mathbf{k}\cdot\mathbf{R}} e^{-i\mathbf{s}\cdot\mathbf{r}_0}\mathbf{F}(\mathbf{k})u_\mathbf{k} \qquad \text{(III.8)}$$

where $\mathbf{F}(\mathbf{k}) \equiv \mathbf{F}(\mathbf{s},\kappa)$ indicates a non–plane-wave mode expansion coefficient, which satisfies

$$\oint_L d\mathbf{u}_{\text{stat}}(\mathbf{r}) = \frac{1}{L^2}\oint_L d\left[\sum_\mathbf{s} e^{i\mathbf{s}\cdot(\mathbf{r}-\mathbf{r}_0)}\mathbf{F}(\mathbf{s},\kappa=0)\right] = -\mathbf{b} \qquad \text{(III.9)}$$

while the scalar mode $u_\mathbf{k}$ satisfies

$$\lim_{\kappa\to 0} u_\mathbf{k} = 1 \qquad \text{(III.10)}$$

Eqs. (III.8)-(III.10) are the basis for a general dislocation theory from mode expansion perspective, prior to any quantization.

At this step, the distinctions between an acoustic phonon expanded from plane waves and a dislocation expanded from localized modes become clear, as summarized in Table 1.

**Table 1.** Comparison between a phonon and a dislocation.

| | Phonon | Dislocation |
|---|---|---|
| Modes | Plane wave | Localized wave |
| Mode expansion | Eq. (III. 2) | Eq. (III. 3) |
| Static limit: expression | $\lim_{\mathbf{k}\to 0}\mathbf{u}_{ph,\mathbf{k}} = 0$ | $\lim_{\kappa\to 0} u_\mathbf{k} = 1$ |
| Static limit: interpretation | Perfect periodic solids | Quenched dislocation |

Before proceeding to a specific type of dislocation, one may be wondering the underlying difference between vector quantization 1) and scalar quantization 2), and whether quantization of a dislocation loop is possible. Briefly speaking, approach 1) will maintain the gauge symmetry of a dislocation [85], i.e. it is a genuine topological defect which cannot be cancelled by local operations. For approach 2), the gauge symmetry is broken (just as a scalar QED breaks photon gauge symmetry [84]), but the displacement field remains finite with dynamic effects retained. Since we are mostly interested in short time scale (~ps) quantum emergent phenomena driven by electron-dislocation and phonon-dislocation interactions, the fate of a dislocation (much longer ~s time scale) is not of concern, validating the approach 2). On the other hand, the quantization of a dislocation loop is possible in theory, which might resemble a quantized closed string in string theory [86]. Yet in practice, this will inevitably cause great mathematical complexity. Therefore, in this study, we only focus on scalar quantization of a straight dislocation line, which retains the mathematical elegancy and can be used to study a wide range of quantum effects associated with dislocations.

Up to now, all the steps are quite generic, without assuming any particular form of $\mathbf{F}(\mathbf{k})$ except for the dislocation's definition in Eq. (III.9). For a particular dislocation with Burgers vector $\mathbf{b}$ and slip plane normal $\mathbf{n}$ within the *xz* plane (Fig. 1), one choice of expansion coefficient $\mathbf{F}(\mathbf{k})$ can be written as [29,76,87]

$$\mathbf{F}_i(\mathbf{k}) = \frac{1}{k_x k^2}\begin{pmatrix} n_i(\mathbf{b}\cdot\mathbf{k}) + b_i(\mathbf{n}\cdot\mathbf{k}) \\ -\frac{1}{(1-\nu)}\frac{k_i(\mathbf{n}\cdot\mathbf{k})(\mathbf{b}\cdot\mathbf{k})}{k^2} \end{pmatrix} \quad \text{(III.11)}$$

in which $i$=$x,y,z$ are the indices of Cartesian direction, $\nu$ is the Poisson ratio. With this choice, it can be directly verified that the corresponding $\mathbf{F}(\mathbf{s}) \equiv \mathbf{F}(\mathbf{k}, \kappa = 0)$ satisfies Eq. (III.9) [76]. For instance, for a screw dislocation line, we have $\mathbf{b} = (0 \quad 0 \quad b)$ and $\mathbf{n} = (0 \quad 1 \quad 0)$, giving $\mathbf{F}_x(\mathbf{s}) = \mathbf{F}_y(\mathbf{s}) = 0$ and $\mathbf{F}_z(\mathbf{s}) = \frac{b}{s^2}\frac{k_y}{k_x}$ in the static limit, and resulting in $\mathbf{u}_z(\mathbf{R}) = \frac{b}{2\pi}\arctan\left(\frac{y}{x}\right)$, which is exactly the classical textbook result of the displacement field of a screw dislocation [1].

The Eqs. (III.8)-(III.10) serve as the basis for the generality of the theory: by assuming different forms of $\mathbf{F}(\mathbf{k})$, the theory can be used to describe dislocations in anisotropic materials; by assuming different boundary conditions of the expansion coefficient, it can even be used to describe different types of localized defects. To proceed, in Section IV, by rewriting $u_\mathbf{k}$ in terms of quantized operators, a quantized displacement of dislocation can be obtained.

## IV. THE QUANTIZATION PROCEDURE OF DISLOCATIONS

To proceed toward a quantized theory, we need to define a canonical coordinate and its conjugate momentum [81]. Apparently, $u_\mathbf{k}$ appearing in Eq. (III.8) seems to be a natural choice of the canonical coordinate. Here, we provide a more general proof of the uniqueness of quantization: putting any prefactor $\Xi_\mathbf{k}$ in front of $u_\mathbf{k}$, and defining the canonical coordinate as $\Xi_\mathbf{k} u_\mathbf{k}$, in the end, it still leads to the same way of quantization.

To see this, we first notice that the classical dislocation's Hamiltonian $H_D$ containing both kinetic energy $T = \frac{\rho}{2}\times \int \sum_{i=1}^{3} \dot{\mathbf{u}}_i^2(\mathbf{R}) d^3\mathbf{R}$ and potential energy $U = \frac{1}{2}\int c_{ijkl} u_{ij} u_{kl} d^3\mathbf{R}$ can be written as

$$H_D = T + U = \frac{1}{2L}\sum_\mathbf{k} T_\mathbf{k} \dot{u}_\mathbf{k} \dot{u}_{-\mathbf{k}} + \frac{1}{2L}\sum_\mathbf{k} W_\mathbf{k} u_\mathbf{k} u_{-\mathbf{k}} \quad \text{(IV.1)}$$

where $\rho$ is the mass density and $c_{ijkl}$ is the stiffness tensor. In an isotropic material, $c_{ijkl} \equiv \lambda\delta_{ij}\delta_{kl} + \mu(\delta_{ik}\delta_{jl} + \delta_{il}\delta_{jk})$ with $\lambda$ and $\mu$ called Lamé parameters. Furthermore $T_\mathbf{k} \equiv \rho|F(\mathbf{k})|^2$ and $W_\mathbf{k} \equiv (\lambda+\mu)[\mathbf{k}\cdot\mathbf{F}(\mathbf{k})]^2 + \mu k^2 |F(\mathbf{k})|^2$, which can be seen by substituting the $c_{ijkl}$ into the classical Hamiltonian.

The canonical momentum conjugate to the canonical coordinate $\Xi_\mathbf{k} u_\mathbf{k}$ can be defined as $p_\mathbf{k} = \frac{\partial \mathcal{L}}{\partial \Xi_\mathbf{k} \dot{u}_\mathbf{k}} = \frac{\partial (T-U)}{\partial \Xi_\mathbf{k} \dot{u}_\mathbf{k}} = \frac{T_\mathbf{k}}{L\Xi_\mathbf{k}}\dot{u}_{-\mathbf{k}}$, where $\mathcal{L} = T - U$ is the Lagrangian, then Eq. (IV.1) can be rewritten in terms of the canonical coordinate $\Xi_\mathbf{k} u_\mathbf{k}$ and its conjugate momentum $p_\mathbf{k}$ as

$$H_D = \frac{1}{2}\sum_\mathbf{k} \frac{p_\mathbf{k} p_{-\mathbf{k}}}{M_\mathbf{k}} + \frac{1}{2L}\sum_\mathbf{k} \frac{W_\mathbf{k}}{\Xi_\mathbf{k}\Xi_{-\mathbf{k}}}(\Xi_\mathbf{k} u_\mathbf{k})(\Xi_{-\mathbf{k}} u_{-\mathbf{k}}) \quad \text{(IV.2)}$$

in which $M_\mathbf{k} \equiv T_\mathbf{k}/(L\Xi_\mathbf{k}\Xi_{-\mathbf{k}})$ plays the role of "mass" in the dislocation's Hamiltonian. As in the case of phonon quantization, now the canonical creation and annihilation operators are defined as

$$\begin{cases} \Xi_\mathbf{k} u_\mathbf{k} = Z_\mathbf{k}\left[a_\mathbf{k} + a_{-\mathbf{k}}^+\right] \\ p_\mathbf{k} = \frac{i\hbar}{2Z_\mathbf{k}}\left[a_\mathbf{k}^+ - a_{-\mathbf{k}}\right] \end{cases} \quad \text{(IV.3)}$$

with $Z_\mathbf{k} \equiv \sqrt{\hbar/2M_\mathbf{k}\Omega_\mathbf{k}}$, $\Omega_\mathbf{k} \equiv \sqrt{W_\mathbf{k}/M_\mathbf{k}\Xi_\mathbf{k}\Xi_{-\mathbf{k}}L} = \sqrt{W_\mathbf{k}/T_\mathbf{k}} = \Omega_{-\mathbf{k}}$. Then we have $\sqrt{\hbar\Xi_\mathbf{k}\Xi_{-\mathbf{k}}L/2T_\mathbf{k}\Omega_\mathbf{k}} = \Xi_\mathbf{k}\sqrt{\hbar L/2T_\mathbf{k}\Omega_\mathbf{k}}$ $Z_\mathbf{k} \equiv \sqrt{\hbar/2M_\mathbf{k}\Omega_\mathbf{k}} = = \Xi_\mathbf{k}\sqrt{\hbar/2m_\mathbf{k}\Omega_\mathbf{k}}$ (where we have assumed $\Xi_\mathbf{k} = \Xi_{-\mathbf{k}}$ and further defined $m_\mathbf{k} \equiv T_\mathbf{k}/L$), and have $Z_\mathbf{k}/\Xi_\mathbf{k} = \sqrt{\hbar/2m_\mathbf{k}\Omega_\mathbf{k}}$ valid in all circumstances. Now we can see that no matter what $\Xi_\mathbf{k}$ we choose, we obtain a unique quantization, since a consistent quantization is only a function of the ratio $Z_\mathbf{k}/\Xi_\mathbf{k}$. In this situation, for simplicity we can choose $\Xi_\mathbf{k} = 1$. Therefore, the displacement field caused by a dislocation Eq. (III.8) can finally be written as in quantized form as

$$\mathbf{u}(\mathbf{R}) = \frac{1}{L^2}\sum_\mathbf{k} e^{i\mathbf{k}\cdot\mathbf{R}} e^{-i\mathbf{s}\cdot\mathbf{r}_0}\mathbf{F}(\mathbf{k})\sqrt{\frac{\hbar}{2m_\mathbf{k}\Omega_\mathbf{k}}}\left[a_\mathbf{k} + a_{-\mathbf{k}}^+\right] \quad \text{(IV.4)}$$

To proceed, we may need to understand the algebraic relation of operators $a_\mathbf{k}$ and $a_\mathbf{k}^+$. A natural choice would be the usual Bosonic commutation $[a_\mathbf{k}, a_{\mathbf{k}'}^+] = \delta_{\mathbf{k}\mathbf{k}'}$; however, we need to bear in mind that any algebraic relation needs to meet the constraint in Eq. (III.10), which indicates that $\lim_{\kappa\to 0} u_\mathbf{k} = \lim_{\kappa\to 0} Z_\mathbf{k}\left[a_\mathbf{k} + a_{-\mathbf{k}}^+\right] = 1 \Rightarrow \lim_{\kappa\to 0}\left[a_\mathbf{k} + a_{-\mathbf{k}}^+\right] = \frac{1}{Z_\mathbf{k}}$.

Therefore, if Bosonic commutation relation $[a_\mathbf{k}, a_{\mathbf{k}'}^+] = \delta_{\mathbf{k}\mathbf{k}'}$ is valid, then we have

$$\begin{aligned} \lim_{\kappa\to 0}\delta_{\mathbf{k}\mathbf{k}'} &= \lim_{\kappa\to 0}[a_{\mathbf{s},\kappa}, a_{\mathbf{s},\kappa}^+] = \lim_{\kappa\to 0}\left[\frac{1}{Z_\mathbf{k}} - a_{-\mathbf{s},-\kappa}^+, \frac{1}{Z_\mathbf{k}} - a_{-\mathbf{s},-\kappa}\right] \\ &= -\lim_{\kappa\to 0}[a_{-\mathbf{s},-\kappa}, a_{-\mathbf{s},-\kappa}^+] = -\lim_{\kappa\to 0}\delta_{\mathbf{k}\mathbf{k}'} \end{aligned}$$

which leads to immediate inconsistency. Therefore, the dislocation's constraint (III.10) leads to a breakdown of the canonical quantization condition.

The breakdown of the canonical commutation relation is not a catastrophe, but can occur in a system with constraints [88]. As a simple example discussed in [89], usual canonical quantization gives $[x, p_x] = i\hbar$ and $[y, p_x] = 0$, which leads to immediate inconsistency if a particle's motion is restricted to a line $x + y = 0$.

## V. THE DISLON HAMILTONIAN

To proceed further, instead of using a canonical quantization condition, we define an alternative quantization condition as

$$[a_{\mathbf{k}}, a^{+}_{\mathbf{k}'}] = \delta_{\mathbf{k}\mathbf{k}'}\,\mathrm{sgn}(\mathbf{k}) \tag{V.1}$$

in which $\mathrm{sgn}(\mathbf{k})$ is a vector sign function satisfying $\mathrm{sgn}(\mathbf{k}) = -\mathrm{sgn}(-\mathbf{k})$, which is elaborated in Appendix A.

Now we can show directly that the operators $(a_{\mathbf{k}}, a^{+}_{\mathbf{k}})$ are consistent with constraint Eq. (III.10):

$$\lim_{\kappa\to 0}\mathrm{sgn}(\mathbf{k}) = \lim_{\kappa\to 0}[a_{\mathbf{k}}, a^{+}_{\mathbf{k}}] = \lim_{\kappa\to 0}[\frac{1}{Z_{\mathbf{k}}} - a^{+}_{-\mathbf{k}}, \frac{1}{Z_{\mathbf{k}}} - a_{-\mathbf{k}}]$$
$$= -\lim_{\kappa\to 0}[a_{-\mathbf{k}}, a^{+}_{-\mathbf{k}}] = -\lim_{\kappa\to 0}\mathrm{sgn}(-\mathbf{k}) = \lim_{\kappa\to 0}\mathrm{sgn}(\mathbf{k})$$

It is worth mentioning that the commutation relation Eq. (V.1) is not the only way leading to quantization – any odd function $\Theta(\mathbf{k})$ satisfying $\Theta(\mathbf{k}) = -\Theta(-\mathbf{k})$ on the right hand side of Eq. (V.1) will be consistent with the constraint Eq. (III.10). However, we can also always absorb $\Theta(\mathbf{k})$ into the normalization factor to rescale operators $(a_{\mathbf{k}}, a^{+}_{\mathbf{k}})$, since the final Hamiltonian should not depend on the choice of $\Theta(\mathbf{k})$. The choice of Eq. (V.1) can then be considered as the simplest choice with a clear physical interpretation as two conventional Bosonic fields, as discussed shortly in Eq. (V.6).

Now substituting Eq. (V.1) back into the classical dislocation's Hamiltonian Eq. (IV.1), the dislon Hamiltonian in 3D can be written as

$$H_D = \sum_{\mathbf{k}} \hbar\Omega(\mathbf{k}) a^{+}_{\mathbf{k}} a_{\mathbf{k}} \tag{V.2}$$

To further simplify the quantized dislocation's Hamiltonian Eq. (V.2), we adopt the concept of a supersymmetric Boson sea to reduce it to a more familiar form [90,91]. Defining $a_{\mathbf{k}1} = a_{\mathbf{k}}$ when $\mathrm{sgn}(\mathbf{k}) > 0$, $a^{+}_{-\mathbf{k}2} = a_{\mathbf{k}}$ when $\mathrm{sgn}(\mathbf{k}) < 0$, we have $a_{-\mathbf{k}2} = a^{+}_{\mathbf{k}}$, thus the operators $a_{\mathbf{k}1}$ and $a_{\mathbf{k}2}$ satisfy normal Boson canonical commutation relation $[a_{\mathbf{k}2}, a^{+}_{\mathbf{k}'2}] = \delta_{\mathbf{k}\mathbf{k}'}$. The dislon Hamiltonian Eq. (V.2) can be rewritten as

$$\begin{aligned} H_D &= \sum_{\mathbf{k}>0} \hbar\Omega(\mathbf{k})\left(a^{+}_{\mathbf{k}_1} a_{\mathbf{k}_1} + \frac{1}{2}\right) \\ &+ \sum_{\mathbf{k}>0} \hbar\Omega(\mathbf{k})\left(a^{+}_{\mathbf{k}_2} a_{\mathbf{k}_2} + \frac{1}{2}\right) \end{aligned} \tag{V.3}$$

in which the $\mathbf{k} \ge 0$ is the shorthand notation of $\mathrm{sgn}(\mathbf{k}) > 0$. In addition, the boundary condition of $u_{\mathbf{k}}$ in Eq. (III.10) is greatly simplified as

$$\begin{aligned} \lim_{\kappa\to 0} u_{\mathbf{k}} &= \lim_{\kappa\to 0}\left(a_{\mathbf{k}1} + a_{\mathbf{k}2}\right) = \lim_{\kappa\to 0}\left(a^{+}_{\mathbf{k}1} + a^{+}_{\mathbf{k}2}\right) \\ &= \lim_{\kappa\to 0} \Xi_{\mathbf{k}} \big/ Z_{\mathbf{k}} = \lim_{\kappa\to 0}\sqrt{\frac{2m_{\mathbf{k}}\Omega_{\mathbf{k}}}{\hbar}} \end{aligned} \tag{V.4}$$

Now performing Keldysh rotation by defining a new set of operators as

$$\begin{aligned} d_{\mathbf{k}} &= \frac{1}{\sqrt{2}}\left(a_{\mathbf{k}1} + a_{\mathbf{k}2}\right), \quad d^{+}_{\mathbf{k}} = \frac{1}{\sqrt{2}}\left(a^{+}_{\mathbf{k}1} + a^{+}_{\mathbf{k}2}\right) \\ f_{\mathbf{k}} &= \frac{1}{\sqrt{2}}\left(-a_{\mathbf{k}1} + a_{\mathbf{k}2}\right), \quad f^{+}_{\mathbf{k}} = \frac{1}{\sqrt{2}}\left(-a^{+}_{\mathbf{k}1} + a^{+}_{\mathbf{k}2}\right) \end{aligned} \tag{V.5}$$

the dislon Hamiltonian Eq. (V.3) can then be rewritten as

$$\begin{aligned} H_D &= \sum_{\mathbf{k}\ge 0} \hbar\Omega_{\mathbf{k}}\left(d^{+}_{\mathbf{k}} d_{\mathbf{k}} + \frac{1}{2}\right) \\ &+ \sum_{\mathbf{k}\ge 0} \hbar\Omega_{\mathbf{k}}\left(f^{+}_{\mathbf{k}} f_{\mathbf{k}} + \frac{1}{2}\right) \end{aligned} \tag{V.6}$$

while the dislon's boundary condition Eq. (V.4) can be rewritten as

$$\lim_{\kappa\to 0} d_{\mathbf{k}} = \lim_{\kappa\to 0} d^{+}_{\mathbf{k}} = \lim_{\kappa\to 0}\sqrt{\frac{m_{\mathbf{k}}\Omega_{\mathbf{k}}}{\hbar}} \equiv C_{\mathrm{s}} \tag{V.7}$$

where $C_{\mathrm{s}}$ is a $\kappa$-independent constant.

With this dislon Hamiltonian Eq. (V.6) and boundary condition Eq. (V.7) in hand, the non-interacting dislon theory in an isotropic medium is complete. In other words: a dislon is composed of *two* independent Bosonic fields, *d*-field and *f*-field, with only the *d*-field constrained by Eq. (V.7), which can be traced to the topological definition of a dislocation.

## VI. THE HAMILTONIAN FOR ELECTON-DISLON INTERACTION

For the electron-dislon interaction, we note that electron density will scatter with the charge imbalance caused by lattice displacement. A generic electron-ion interaction can be written in terms of deformation potential scattering as [81]

$$H_{e-ion} = \int d^3\mathbf{R}\rho_e(\mathbf{R})\sum_{j=1}^{N}\nabla_{\mathbf{R}}V_{ei}(\mathbf{R}-\mathbf{R}^0_j)\cdot\mathbf{u}(\mathbf{R}^0_j) \tag{VI.1}$$

where the summation over $j$ is over $N$ atoms in the solids, $\mathbf{R}^0_j$ is the atomic coordinate for the $j^{\mathrm{th}}$ atom in a perfect crystal, and the electron-ion Coulomb potential can be expanded as $V_{ei}(\mathbf{R}-\mathbf{R}_j^0)=\frac{1}{L^3}\sum_{\mathbf{k}}V_{\mathbf{k}}e^{i\mathbf{k}\cdot(\mathbf{R}-\mathbf{R}_j^0)}$, with Fourier component $V_{\mathbf{k}}=\frac{4\pi Ze}{k^2+k_{TF}^2}=V_{-\mathbf{k}}$ ( $k_{TF}$ is Thomas-Fermi screening), and the charge density operator gives $\rho_e(\mathbf{R})=$. $\frac{e}{L^3}\sum_{\mathbf{kp}\sigma}e^{-i\mathbf{p}\cdot\mathbf{R}}c_{\mathbf{k}+\mathbf{p}\sigma}^{+}c_{\mathbf{k}\sigma}$ , where $c^+$ and $c$ are the electron creation and annihilation operators, respectively.

For the specific case of the displacement field caused by the dislocation, we could substitute Eq. (IV.4) and the Coulomb potential expansion into Eq. (VI.1), where we first obtain

$$\sum_{j=1}^{N}\nabla_{\mathbf{R}}V_{ei}(\mathbf{R}-\mathbf{R}_j^0)\cdot\mathbf{u}_j$$

$$=\frac{N}{L^5}\sum_{\mathbf{k}}V_{\mathbf{k}}e^{i\mathbf{k}\cdot\mathbf{R}}i\mathbf{k}\cdot\mathbf{F}(\mathbf{k})\sqrt{\frac{\hbar}{2m_{\mathbf{k}}\Omega_{\mathbf{k}}}}\left[a_{\mathbf{k}}+a_{-\mathbf{k}}^{+}\right]$$

Then, using Eq. (V.5), the classical Hamiltonian Eq. (VI.1) for an electron scattering from a single dislocation with core located at $\mathbf{r}_0$ can then be rewritten in a quantized way as

$$\begin{aligned}H_{e-dis}=&\sum_{\substack{\mathbf{k}'\sigma\\ \mathbf{k}\geq 0}}\sqrt{2}g_{\mathbf{k}}e^{-i\mathbf{s}\cdot\mathbf{r}_0}c_{\mathbf{k}'+\mathbf{k}\sigma}^{+}c_{\mathbf{k}'\sigma}d_{\mathbf{k}}\\&+\sum_{\substack{\mathbf{k}'\sigma\\ \mathbf{k}\geq 0}}\sqrt{2}g_{\mathbf{k}}^{*}e^{+i\mathbf{s}\cdot\mathbf{r}_0}c_{\mathbf{k}'-\mathbf{k}\sigma}^{+}c_{\mathbf{k}'\sigma}d_{\mathbf{k}}^{+}\end{aligned}\qquad\text{(VI.2)}$$

where the electron-dislon coupling constant gives $g_{\mathbf{k}}\equiv\frac{eN}{L^5}\times V_{\mathbf{k}}\left[i\mathbf{k}\cdot\mathbf{F}(\mathbf{k})\right]\sqrt{\frac{\hbar}{2m_{\mathbf{k}}\Omega_{\mathbf{k}}}}=g_{-\mathbf{k}}^{*}$. Here we see that the deformation potential scattering only couples with the displacement-like *d*-field of a dislon, but not momentum-like *f*-field.

Up to this step, we have only treated one single dislocation line. To generalize to multiple dislocations, we can assume a set of dislocation core locations $\{\mathbf{r}_j\}$ and sum over all dislocations,; however, later it can be shown that it would be much more convenient if we add multiple dislocations in effective theories, instead of adding into the interaction at this moment.

## VII. THE PHONON-DISLON FLUTTERING

Other than the case for electron-dislon scattering, which is intrinsically deformation potential scattering as shown in Eq. (VI.2), the dominant interaction for phonon-dislon interaction is the so-called fluttering mechanism, which is a drag-like scattering coupling the first-order time-derivative of the phonon and the dislocation displacement field [29,32,87]. Previously we derived the phonon-dislocation fluttering mechanism using 1D dislocation quantization [36]; here we present a full 3D quantization approach. As a brief review of phonon quantization, the lattice displacement of a phonon $\mathbf{u}_{ph}(\mathbf{R})$ with a single phonon mode can be written as [83]

$$\mathbf{u}_{ph}(\mathbf{R})=\frac{1}{L^{3/2}}\sum_{\mathbf{k}}\sqrt{\frac{\hbar}{2\rho\omega_{\mathbf{k}}}}\left(b_{\mathbf{k}}+b_{-\mathbf{k}}^{+}\right)\boldsymbol{\varepsilon}_{\mathbf{k}}e^{i\mathbf{k}\cdot\mathbf{R}}\qquad\text{(VII.1)}$$

where $\omega_{\mathbf{k}}$ is the phonon dispersion, $L^3$ is the system volume, and $\rho$ is the mass density. The hermiticity of $\mathbf{u}$ requires that the polarization vector satisfies $\boldsymbol{\varepsilon}_{\mathbf{k}}=\boldsymbol{\varepsilon}_{-\mathbf{k}}^{*}$.

Now defining the canonical conjugate momentum density as $\mathbf{p}_{ph}\equiv\frac{\partial\mathcal{L}}{\partial\dot{\mathbf{u}}_{ph}}=\rho\dot{\mathbf{u}}_{ph}$, we have

$$\mathbf{p}_{ph}(\mathbf{R})=\frac{1}{\sqrt{L^3}}\sum_{\mathbf{k}}i\sqrt{\frac{\rho\hbar\omega_{\mathbf{k}}}{2}}\left(-b_{\mathbf{k}}+b_{-\mathbf{k}}^{+}\right)\boldsymbol{\varepsilon}_{\mathbf{k}}e^{i\mathbf{k}\cdot\mathbf{R}}\qquad\text{(VII.2)}$$

It can be verified directly that $[b_{\mathbf{k}},b_{\mathbf{k}'}^{+}]=\delta_{\mathbf{kk}'}$ is consistent with the real-space canonical quantization condition $[\mathbf{u}_{ph}(\mathbf{R}),\mathbf{p}_{ph}(\mathbf{R}')]=+i\hbar\delta^{(3)}(\mathbf{R}-\mathbf{R}')$ using the facts that $\sum_{\mathbf{k}}e^{i\mathbf{k}\cdot(\mathbf{R}-\mathbf{R}')}=L^3\delta^{(3)}(\mathbf{R}-\mathbf{R}')$ and $\int d^3\mathbf{R}e^{i(\mathbf{k}-\mathbf{k}')\cdot\mathbf{R}}=L^3\delta_{\mathbf{k},\mathbf{k}'}$.

Now if we further define the displacement and momentum operators in Fourier transformed space

$$\begin{aligned}\mathbf{u}_{ph}(\mathbf{R})&=\sum_{\mathbf{k}}\mathbf{u}_{ph,\mathbf{k}}\frac{e^{i\mathbf{k}\cdot\mathbf{R}}}{L^{3/2}};\ \mathbf{u}_{ph,\mathbf{k}}=\int\mathbf{u}_{ph}(\mathbf{R})\frac{e^{-i\mathbf{k}\cdot\mathbf{R}}}{L^{3/2}}d^3\mathbf{R}\\ \mathbf{p}_{ph}(\mathbf{R})&=\sum_{\mathbf{k}}\mathbf{p}_{ph,\mathbf{k}}\frac{e^{i\mathbf{k}\cdot\mathbf{R}}}{L^{3/2}};\ \mathbf{p}_{ph,\mathbf{k}}=\int\mathbf{p}_{ph}(\mathbf{R})\frac{e^{-i\mathbf{k}\cdot\mathbf{R}}}{L^{3/2}}d^3\mathbf{R}\end{aligned}\qquad\text{(VII.3)}$$

then we have

$$\begin{aligned}\mathbf{u}_{ph,\mathbf{k}}&=\sqrt{\frac{\hbar}{2\rho\omega_{\mathbf{k}}}}\left(b_{\mathbf{k}}+b_{-\mathbf{k}}^{+}\right)\boldsymbol{\varepsilon}_{\mathbf{k}}\\ \mathbf{p}_{ph,\mathbf{k}}&=i\sqrt{\frac{\rho\hbar\omega_{\mathbf{k}}}{2}}\left(-b_{\mathbf{k}}+b_{-\mathbf{k}}^{+}\right)\boldsymbol{\varepsilon}_{\mathbf{k}}\end{aligned}\qquad\text{(VII.4)}$$

which satisfy $\mathbf{u}_{ph,-\mathbf{k}}=\mathbf{u}_{ph,\mathbf{k}}^{*}$ and $\mathbf{p}_{ph,-\mathbf{k}}=\mathbf{p}_{ph,\mathbf{k}}^{*}$ and $[\mathbf{u}_{ph,\mathbf{k}},\mathbf{p}_{ph,\mathbf{k}'}]=i\hbar\delta_{\mathbf{k},-\mathbf{k}'}$.

The phonon Hamiltonian can be written as

$$H_{ph}=\sum_{\mathbf{k}}\hbar\omega_{\mathbf{k}}\left(b_{\mathbf{k}}^{+}b_{\mathbf{k}}+\frac{1}{2}\right)\qquad\text{(VII.5)}$$

The phonon-dislocation interaction Hamiltonian is given by the following expression, called a fluttering mechanism, which is the cross term between the phonon $\dot{\mathbf{u}}_{ph}(\mathbf{R})$ and dislon $\dot{\mathbf{u}}(\mathbf{R})$ appearing in the kinetic energy *T*. Since $T=\frac{1}{2}\rho\int\dot{\mathbf{u}}_{tot}^2(\mathbf{R})d^3\mathbf{R}$ and $\dot{\mathbf{u}}_{tot}(\mathbf{R})=\dot{\mathbf{u}}_{ph}(\mathbf{R})+\dot{\mathbf{u}}(\mathbf{R})$ , we have [29,87]

$$H_{flu}=\rho\int\dot{\mathbf{u}}_{ph}(\mathbf{R})\cdot\dot{\mathbf{u}}(\mathbf{R})d^3\mathbf{R}=\int\mathbf{p}_{ph}(\mathbf{R})\cdot\dot{\mathbf{u}}(\mathbf{R})d^3\mathbf{R}\qquad\text{(VII.6)}$$

Now substituting Eq. (VII.1) and Eq. (IV.3) back into Eq. (VII.6), using the fact that $p_{\mathbf{k}} = \frac{i}{\Xi_{\mathbf{k}}}\sqrt{\frac{\hbar m_{\mathbf{k}}\Omega_{\mathbf{k}}}{2}}\left[a_{\mathbf{k}}^{+} - a_{-\mathbf{k}}\right]$ $= \frac{m_{\mathbf{k}}}{\Xi_{\mathbf{k}}}\dot{u}_{-\mathbf{k}}$ , and recalling Eq. (V.5), we finally obtain the quantized phonon-dislon fluttering Hamiltonian as

$$\begin{aligned}H_{flu} = +\frac{\hbar}{L^{1/2}}\sum_{\mathbf{k}>0}\sqrt{\frac{\rho\omega_{\mathbf{k}}\Omega_{\mathbf{k}}}{2m_{\mathbf{k}}}}\left[\boldsymbol{\varepsilon}_{\mathbf{k}}\cdot\mathbf{F}(\mathbf{k})\right]\times \\ \left[e^{+is\cdot\mathbf{r}_0}\left(-b_{\mathbf{k}}+b_{-\mathbf{k}}^{+}\right)f_{\mathbf{k}}^{+} - e^{-is\cdot\mathbf{r}_0}\left(-b_{-\mathbf{k}}+b_{\mathbf{k}}^{+}\right)f_{\mathbf{k}}\right]\end{aligned} \quad \text{(VII.7)}$$

where we have used the facts that $\mathbf{F}(\mathbf{k}) = \mathbf{F}(-\mathbf{k})$, $\Omega_{-\mathbf{k}} = \Omega_{\mathbf{k}}$ , $\omega_{-\mathbf{k}} = \omega_{\mathbf{k}}$, and $m_{-\mathbf{k}} = m_{\mathbf{k}}$.

In other words, we have reached a nontrivial conclusion about the dislon: the non-interacting dislon theory Eq. (V.6) is composed of two independent fields *d* and *f*, where the coupling with an electron is through the displacement-like *d*-field with constraint Eq. (V.7), while the coupling with a phonon through fluttering is through the momentum-like *f*-field without constraint. Such decoupling shows great advantage when constructing an effective theory.

## VIII. PHONON-DISLON ANHARMONICTY

Given the importance of anharmonic interactions to thermal transport, in this section, we derive a framework to study how the phonon-dislon anharmonic interaction may change the phonon properties. We notice that the integral of the cross term of strain energy $\mathbf{u}\cdot\mathbf{u}_{ph}$ is zero according to classical theory [29,87,92]. Therefore, we only need to consider the 3rd order anharmonic term. The generic third order anharmonic interaction of the displacement field can be written as [93]

$$\begin{aligned}H_{Anh} = \frac{1}{6}\iiint \left.\frac{\partial^3 U}{\partial \mathbf{u}_a^{tot}(\mathbf{R}_1)\partial \mathbf{u}_b^{tot}(\mathbf{R}_2)\partial \mathbf{u}_c^{tot}(\mathbf{R}_3)}\right|_{\mathbf{u}=0} \\ \times \mathbf{u}_a^{tot}(\mathbf{R}_1)\mathbf{u}_b^{tot}(\mathbf{R}_2)\mathbf{u}_c^{tot}(\mathbf{R}_3)d^3\mathbf{R}_1 d^3\mathbf{R}_2 d^3\mathbf{R}_3\end{aligned} \quad \text{(VIII.1)}$$

in which $a,b,c = 1,2,3$ are the Cartesian coordinate components, and Einstein's summation convention has been adopted.

Now we note that the total displacement $\mathbf{u}^{tot} = \mathbf{u} + \mathbf{u}_{ph}$ is a sum over both the phonon contribution $\mathbf{u}_{ph}$ and dislocation contribution $\mathbf{u}$ , enabling us to re-arrange Eq. (VIII.1) according to the order of the dislocation's displacement $\mathbf{u}$:

0th order in $\mathbf{u}$: $H_{Anh} \sim \mathbf{u}_{ph}^3$ , which gives the usual 3rd order three-phonon anharmonicity but does not involve any interaction with the dislon, hence will not be considered.

1st order in $\mathbf{u}$: $H_{Anh} \sim \mathbf{u}^1\mathbf{u}_{ph}^2$ which gives the scenario that one incoming phonon interacts with one dislon and the phonon gets scattered. This is the most commonly encountered scenario given the high phonon density of states.

2nd order in $\mathbf{u}$: $H_{Anh} \sim \mathbf{u}^2\mathbf{u}_{ph}^1$ which gives the scenario that one dislon interacts with a phonon and gets scattered, which happens much less frequently since the dislon density of states is much lower. For this reason, we will not consider this scenario either.

3rd order in $\mathbf{u}$: $H_{Anh} \sim \mathbf{u}^3$ which is not only very rare (one dislon is split into two or vice versa) but also doesn't involve any interaction with phonons. We will not consider this scenario either.

In this sense, the only interaction term of interest is thus the 1st order in $\mathbf{u}$:

$$\begin{aligned}H_{Anh} &= \frac{1}{2}\iiint \left.\frac{\partial^3 U}{\partial \mathbf{u}_a(\mathbf{R}_1)\partial \mathbf{u}_b(\mathbf{R}_2)\partial \mathbf{u}_c(\mathbf{R}_3)}\right|_{\mathbf{u}=0} \\ &\quad \mathbf{u}_{ph,a}(\mathbf{R}_1)\mathbf{u}_{ph,b}(\mathbf{R}_2)\mathbf{u}_c(\mathbf{R}_3)d^3\mathbf{R}_1 d^3\mathbf{R}_2 d^3\mathbf{R}_3 \\ &= \frac{1}{2L^5}\sum_{\mathbf{k}_1\mathbf{k}_2\mathbf{k}_3}\iiint \begin{matrix}\left.\frac{\partial^3 U}{\partial \mathbf{u}_a(\mathbf{R}_1)\partial \mathbf{u}_b(\mathbf{R}_2)\partial \mathbf{u}_c(\mathbf{R}_3)}\right|_{\mathbf{u}=0} \\ \times e^{i\mathbf{k}_2\cdot(\mathbf{R}_2-\mathbf{R}_1)}e^{i\mathbf{k}_3\cdot(\mathbf{R}_3-\mathbf{R}_1)}e^{i(\mathbf{k}_1+\mathbf{k}_2+\mathbf{k}_3)\cdot\mathbf{R}_1}e^{-is_3\cdot\mathbf{r}_0}\end{matrix} \\ &\quad \times \mathbf{u}_{ph,\mathbf{k}_1 a}\mathbf{u}_{ph,\mathbf{k}_2 b}\mathbf{F}_c(\mathbf{k}_3)u_{\mathbf{k}_3}d^3\mathbf{R}_1 d^3\mathbf{R}_2 d^3\mathbf{R}_3\end{aligned}$$

Since we are considering a translationally invariant system (large size of solid), the choice of origin does not matter (i.e. independent of $\mathbf{R}_1$ ), but only the relative distance between two spatial points matters (i.e. $\mathbf{R}_2 - \mathbf{R}_1$ and $\mathbf{R}_3 - \mathbf{R}_1$), hence we could define the coefficient $V_{\mathbf{k}_1\mathbf{k}_2\mathbf{k}_3}^{abc}$ as

$$\begin{aligned}V_{\mathbf{k}_1\mathbf{k}_2\mathbf{k}_3}^{abc} \equiv \iint \left.\frac{\partial^3 U}{\partial \mathbf{u}_a(\mathbf{R}_1)\partial \mathbf{u}_b(\mathbf{R}_2)\partial \mathbf{u}_c(\mathbf{R}_3)}\right|_{\mathbf{u}=0} e^{i\mathbf{k}_2\cdot(\mathbf{R}_2-\mathbf{R}_1)} \\ \times e^{i\mathbf{k}_3\cdot(\mathbf{R}_3-\mathbf{R}_1)}d^3\left(\mathbf{R}_2-\mathbf{R}_1\right)d^3\left(\mathbf{R}_3-\mathbf{R}_1\right)\end{aligned} \quad \text{(VIII.2)}$$

This enables the integration over $\mathbf{R}_1$ independently. Therefore, the anharmonic phonon-dislon interaction Hamiltonian can further be rewritten as

$$H_{Anh} = \frac{1}{2L^2}\sum_{\mathbf{k}_1\mathbf{k}_2\mathbf{k}_3}\begin{matrix}\delta^{(3)}(\mathbf{k}_1+\mathbf{k}_2+\mathbf{k}_3)V_{\mathbf{k}_1\mathbf{k}_2\mathbf{k}_3}^{abc} \\ \times e^{-is_3\cdot\mathbf{r}_0}\mathbf{F}_c(\mathbf{k}_3)\mathbf{u}_{ph,\mathbf{k}_1 a}\mathbf{u}_{ph,\mathbf{k}_2 b}u_{\mathbf{k}_3}\end{matrix} \quad \text{(VIII.3)}$$

in which $\delta^{(3)}(\mathbf{k}_1+\mathbf{k}_2+\mathbf{k}_3)$ is a discrete Kronecker-delta function.

Substituting Eqs. (VII.4) and (IV.3) into Eq. (VIII.3), we obtain the quantized phonon-dislon anharmonic interaction Hamiltonian

$$\begin{aligned}H_{Anh} = \frac{\hbar}{4\sqrt{2}\rho L^2}\sum_{\mathbf{k}_1\mathbf{k}_2\mathbf{k}_3} A(\mathbf{k}_1,\mathbf{k}_2,\mathbf{k}_3)e^{-is_3\cdot\mathbf{r}_0} \\ \times\left(b_{\mathbf{k}_1}+b_{-\mathbf{k}_1}^{+}\right)\left(b_{\mathbf{k}_2}+b_{-\mathbf{k}_2}^{+}\right)\left(a_{\mathbf{k}_3}+a_{-\mathbf{k}_3}^{+}\right)\end{aligned} \quad \text{(VIII.4)}$$

where $A(\mathbf{k}_1,\mathbf{k}_2,\mathbf{k}_3)$ is the anharmonic phonon-dislon coupling strength, which is defined as

$$A(\mathbf{k}_1,\mathbf{k}_2,\mathbf{k}_3) \equiv \delta^{(3)}(\mathbf{k}_1+\mathbf{k}_2+\mathbf{k}_3)V^{abc}_{\mathbf{k}_1\mathbf{k}_2\mathbf{k}_3} \times \boldsymbol{\varepsilon}^a_{\mathbf{k}_1}\boldsymbol{\varepsilon}^b_{\mathbf{k}_2}\mathbf{F}_c(\mathbf{k}_3)\sqrt{\frac{\hbar}{m_{\mathbf{k}_3}\Omega_{\mathbf{k}_3}\omega_{\mathbf{k}_1}\omega_{\mathbf{k}_2}}} \quad \text{(VIII.5)}$$

Now using Eq. (V.5) , we rewrite Eq. (VIII.4) as

$$H_{Anh} = \frac{\hbar}{4\rho L^2}\sum_{\substack{\mathbf{k}_1\mathbf{k}_2\\ \mathbf{k}_3>0}}\left(b_{\mathbf{k}_1}+b^+_{-\mathbf{k}_1}\right)\left(b_{\mathbf{k}_2}+b^+_{-\mathbf{k}_2}\right)\times \left[e^{-i\mathbf{s}_3\cdot\mathbf{r}_0}A(\mathbf{k}_1,\mathbf{k}_2,\mathbf{k}_3)d_{\mathbf{k}_3}+e^{+i\mathbf{s}_3\cdot\mathbf{r}_0}A(\mathbf{k}_1,\mathbf{k}_2,-\mathbf{k}_3)d^+_{\mathbf{k}_3}\right] \quad \text{(VIII.6)}$$

which concludes the anharmonic phonon-dislon interaction Hamiltonian.

## IX. ACTION FORMS OF THE ELECTRON-PHONON-DISLON INTERACTING SYSTEM

The constraint on the dislon's *d*-field Eq. (V.7) causes great technical difficulties in using a canonical operator approach. To solve a system with constrained dynamics and highlight the influence of a dislon on purely electronic or phononic degrees of freedom, a functional integral approach is adopted [88]. The first step is to rewrite all the Hamiltonians above into action form.

The transformation of Fermion (F) and Boson (B) operators between imaginary time and Matsubara frequency can be written as [81]

$$\begin{aligned} F_{\mathbf{k}}(\tau) &= \sum_n F_{\mathbf{k}n}\frac{e^{+ip_n\tau}}{\sqrt{\beta}},\ F_{\mathbf{k}n} = \int_0^\beta F_{\mathbf{k}}(\tau)\frac{e^{-ip_n\tau}}{\sqrt{\beta}}d\tau \\ B_{\mathbf{k}}(\tau) &= \sum_n B_{\mathbf{k}n}\frac{e^{+i\omega_n\tau}}{\sqrt{\beta}},\ B_{\mathbf{k}n} = \int_0^\beta B_{\mathbf{k}}(\tau)\frac{e^{-i\omega_n\tau}}{\sqrt{\beta}}d\tau \end{aligned} \quad \text{(IX.1)}$$

in which $p_n \equiv (2n+1)\pi/\beta$ and $\omega_n \equiv 2n\pi/\beta$ are the Fermionic and Bosonic Matsubara frequency, respectively, and $\beta = 1/k_BT$ is the inverse temperature.

The non-interacting dislon Hamiltonian Eq. (V.6) can be rewritten in terms of action form in the Matsubara frequency domain as

$$\begin{aligned} S_{dis}[\bar{d},d,\bar{f},f] = &\sum_{n,\mathbf{k}\ge 0}\bar{d}_{\mathbf{k}n}(-i\omega_n+\hbar\Omega_{\mathbf{k}})d_{\mathbf{k}n} \\ &+\sum_{n,\mathbf{k}\ge 0}\bar{f}_{\mathbf{k}n}(-i\omega_n+\hbar\Omega_{\mathbf{k}})f_{\mathbf{k}n} \end{aligned} \quad \text{(IX.2)}$$

where $\bar{d},d,\bar{f},f$ are the dislon fields written in coherent state form and $n$=0, ±1, ±2….are the Matsubara indices.

The electron action written in the functional integral form can be written as

$$S_e[\bar{\psi},\psi] = \sum_{n\mathbf{p}\sigma}\bar{\psi}_{\mathbf{p}n\sigma}(-ip_n+E_{\mathbf{p}}-\mu)\psi_{\mathbf{p}n\sigma} \quad \text{(IX.3)}$$

where $E_{\mathbf{p}}$ is the electron single-particle energy, $\mu$ is the chemical potential, and $\bar{\psi},\psi$ are the electron fields.

Now defining momentum-space charge density as $\rho_{n\mathbf{k}} \equiv \sum_{\mathbf{k}'m\sigma}\bar{\psi}_{m+n,\mathbf{k}'+\mathbf{k}\sigma}\psi_{m\mathbf{k}'\sigma}$ , the electron-dislon interaction Hamiltonian Eq. (VI.2) can be rewritten as an action form

$$\begin{aligned} S_{e-dis}[\bar{\psi},\psi,\bar{d},d] = &\sum_{n,\mathbf{k}\ge 0}\sqrt{\frac{2}{\beta}}g_{\mathbf{k}}e^{-i\mathbf{s}\cdot\mathbf{r}_0}\rho_{n\mathbf{k}}d_{n\mathbf{k}} \\ &+\sum_{n,\mathbf{k}\ge 0}\sqrt{\frac{2}{\beta}}g^*_{\mathbf{k}}e^{+i\mathbf{s}\cdot\mathbf{r}_0}\rho_{-n-\mathbf{k}}\bar{d}_{n\mathbf{k}} \end{aligned} \quad \text{(IX.4)}$$

where $g_{\mathbf{k}} \equiv \frac{eN}{L^5}V_{\mathbf{k}}\left[i\mathbf{k}\cdot\mathbf{F}(\mathbf{k})\right]\sqrt{\frac{\hbar}{2m_{\mathbf{k}}\Omega_{\mathbf{k}}}} = g^*_{-\mathbf{k}}$ , and boundary condition Eq. (V.7) is now rewritten from a canonical operator to a coherent state in the Matsubara frequency domain as

$$\lim_{\kappa\to 0}d_{\mathbf{k}} = C_{\mathbf{s}} \Rightarrow \lim_{\kappa\to 0}d_{\mathbf{k}}(\tau) = C_{\mathbf{s}} \Rightarrow \lim_{\kappa\to 0}d_{\mathbf{k}}(\tau)e^{-i\omega_n\tau} = C_{\mathbf{s}}e^{-i\omega_n\tau} \Rightarrow$$

$$\lim_{\kappa\to 0}\frac{1}{\sqrt{\beta}}\int_0^\beta d_{\mathbf{k}}(\tau)e^{-i\omega_n\tau}d\tau = C_{\mathbf{s}}\frac{1}{\sqrt{\beta}}\int_0^\beta e^{-i\omega_n\tau}d\tau \Rightarrow \lim_{\kappa\to 0}d_{\mathbf{k}n} = \sqrt{\beta}C_{\mathbf{s}}$$

from which we have

$$\begin{aligned} \lim_{\kappa\to 0}d_{\mathbf{k}} &= \lim_{\kappa\to 0}d_{\mathbf{s},\kappa} \equiv d_{\mathbf{s}0} = \lim_{\kappa\to 0}\sqrt{\frac{\beta m_{\mathbf{k}}\Omega_{\mathbf{k}}}{\hbar}} \equiv \sqrt{\beta}C_{\mathbf{s}} \\ \lim_{\kappa\to 0}\bar{d}_{\mathbf{k}} &= \lim_{\kappa\to 0}\bar{d}_{\mathbf{s},\kappa} \equiv \bar{d}_{\mathbf{s}0} = \lim_{\kappa\to 0}\sqrt{\frac{\beta m_{\mathbf{k}}\Omega_{\mathbf{k}}}{\hbar}} = \sqrt{\beta}C_{\mathbf{s}} \end{aligned} \quad \text{(IX.5)}$$

where $C_{\mathbf{s}}$ is an **s**-dependent constant defined in Eq. (V.7), and $\mathrm{sgn}(\mathbf{s}) \ge 0$ (Appendix A).

The phonon action can be written into action form as

$$S_{ph}[\bar{b},b] = \sum_{n\mathbf{k}}\bar{b}_{\mathbf{k}n}(-i\omega_n+\hbar\omega_{\mathbf{k}})b_{\mathbf{k}n} \quad \text{(IX.6)}$$

in which $\bar{b},b$ is the phonon field written in Bosonic coherent state form, and $\omega_{\mathbf{k}}$ is the phonon dispersion relation. The phonon-dislon fluttering interaction Hamiltonian Eq. (VII.7) can be rewritten in terms of action form as

$$S_{flu} = \hbar\sum_{\mathbf{k}>0,n}\sqrt{\frac{\rho\omega_{\mathbf{k}}\Omega_{\mathbf{k}}}{2T_{\mathbf{k}}}}\left[\boldsymbol{\varepsilon}_{\mathbf{k}}\cdot\mathbf{F}(\mathbf{k})\right]\times \left[e^{+i\mathbf{s}\cdot\mathbf{r}_0}\left(-b_{\mathbf{k}n}+\bar{b}_{-\mathbf{k},-n}\right)\bar{f}_{\mathbf{k}n}-e^{-i\mathbf{s}\cdot\mathbf{r}_0}\left(-b_{-\mathbf{k},-n}+\bar{b}_{\mathbf{k}n}\right)f_{\mathbf{k}n}\right] \quad \text{(IX.7)}$$

Now we rewrite the anharmonic Hamiltonian Eq. (VIII.6) into action form in imaginary time directly as

$$S_{Anh} = \frac{\hbar}{4\rho L^2}\int_0^\beta d\tau\sum_{\substack{\mathbf{k}_1\mathbf{k}_2\\ \mathbf{k}_3>0}}\left(b_{\mathbf{k}_1}(\tau)+\bar{b}_{-\mathbf{k}_1}(\tau)\right)\left(b_{\mathbf{k}_2}(\tau)+\bar{b}_{-\mathbf{k}_2}(\tau)\right)\times \left[e^{-i\mathbf{s}_3\cdot\mathbf{r}_0}A(\mathbf{k}_1,\mathbf{k}_2,\mathbf{k}_3)d_{\mathbf{k}_3}(\tau)+e^{i\mathbf{s}_3\cdot\mathbf{r}_0}A(\mathbf{k}_1,\mathbf{k}_2,-\mathbf{k}_3)\bar{d}_{\mathbf{k}_3}(\tau)\right]$$

Now using the Keldysh rotated field defined in Eq. (XI.3), i.e. $\phi_{\mathbf{k}n} = \frac{1}{\sqrt{2}}\left(b_{\mathbf{k}n}+\bar{b}_{-\mathbf{k}-n}\right)$ which has the physical meaning of a phonon displacement field, and using Eq. (IX.1) , we obtain the anharmonic phonon-dislon action in the Matsubara frequency domain as

$$S_{Anh}[\phi,\bar{d},d]=\frac{\hbar}{2\rho\sqrt{\beta}L^{2}}\sum_{\substack{\mathbf{k}_1\mathbf{k}_2\\ \mathbf{k}_3>0\\ m,n,p}}\phi_{1\mathbf{k}_1,m}\phi_{1\mathbf{k}_2,n}\times\left[\begin{array}{l} e^{-i\mathbf{s}_3\cdot\mathbf{r}_0}A(\mathbf{k}_1,\mathbf{k}_2,\mathbf{k}_3)\delta_{m+n,-p}d_{\mathbf{k}_3,p}\\ +e^{i\mathbf{s}_3\cdot\mathbf{r}_0}A(\mathbf{k}_1,\mathbf{k}_2,-\mathbf{k}_3)\delta_{m+n,p}\bar{d}_{\mathbf{k}_3,p}\end{array}\right] \qquad \text{(IX.8)}$$

where the summation *m,n,p* is over the Bosonic Matsubara frequency.

With Eqs. (IX.2)-(IX.8) in hand, we are ready to derive the electron and phonon effective theories when they start to interact with a dislon.

## X. EFFECTIVE ELECTRON THEORY

Taking into account the constraint Eq. (IX.5), we define an effective action of the electron $S_{eff}[\bar{\psi},\psi]$ by integrating over the dislon degrees of freedom as

$$e^{-S_{eff}[\bar{\psi},\psi]}\equiv e^{-S_e[\bar{\psi},\psi]}\int D[\bar{d},d]e^{-S_{dis}[\bar{d},d]-S_{e-dis}[\bar{\psi},\psi,\bar{d},d]}\times\prod_{n\mathbf{s}\geq 0}\delta(d_{n\mathbf{s}0}-\sqrt{\beta}C_{\mathbf{s}})\delta(\bar{d}_{n\mathbf{s}0}-\sqrt{\beta}C_{\mathbf{s}}) \qquad \text{(X.1)}$$

in which $S_D[\bar{d},d]=\sum_{n,\mathbf{k}\geq 0}\bar{d}_{\mathbf{k}n}(-i\omega_n+\hbar\Omega_{\mathbf{k}})d_{\mathbf{k}n}$ is the *d*-field portion of the dislon field, since only the *d*-field of the dislon couples with electrons.

The method of imposing the constraint of Eq. (IX.5) onto Eq. (X.1) is to introduce a Dirac δ-function, which is similar to the gauge-fixing method used in the Faddeev-Popov ghost [94], or to the length-fixing method in the nonlinear sigma model [95], while here it is fixing the dislon boundary condition instead of gauge fixing or length fixing.

Performing the Fourier transform so that $\delta(d_{n\mathbf{s}0}-\sqrt{\beta}C_{\mathbf{s}})=\int\frac{d\bar{k}_{n\mathbf{s}}}{2\pi}e^{i\bar{k}_{n\mathbf{s}}(d_{n\mathbf{s}0}-\sqrt{\beta}C_{\mathbf{s}})}$, Eq. (X.1) can be rewritten as

$$e^{-S_{eff}[\bar{\psi},\psi]}=e^{-S_e[\bar{\psi},\psi]}\times\int D[\bar{d},d]D[\bar{k},k]e^{-S_{dis}[\bar{d},d]-S_{e-dis}[\bar{\psi},\psi,\bar{d},d]}\times e^{i\sum_{n,\mathbf{s}\geq 0}\left(\bar{k}_{n\mathbf{s}}d_{n\mathbf{s}0}+k_{n\mathbf{s}}\bar{d}_{n\mathbf{s}0}-\sqrt{\beta}C_{\mathbf{s}}k_{n\mathbf{s}}-\sqrt{\beta}C_{\mathbf{s}}\bar{k}_{n\mathbf{s}}\right)}$$

where we have defined a functional measure $D[\bar{k},k]\equiv\prod_{n\mathbf{s}}\int\frac{d\bar{k}_{n\mathbf{s}}}{2\pi}\frac{dk_{n\mathbf{s}}}{2\pi}$.

Now integrating over the dislon degrees of freedom and noticing $\sum_{n,\mathbf{s}\geq 0}\left(\bar{k}_{n\mathbf{s}}d_{n\mathbf{s}0}+k_{n\mathbf{s}}\bar{d}_{n\mathbf{s}0}\right)=\sum_{n,\mathbf{k}\geq 0}\left(\bar{k}_{n\mathbf{s}}\delta_{\kappa 0}d_{n\mathbf{k}}+k_{n\mathbf{s}}\delta_{\kappa 0}\bar{d}_{n\mathbf{k}}\right)$, we obtain

$$\int D[\bar{d},d]e^{-S_{dis}[\bar{d},d]-S_{e-dis}[\bar{\psi},\psi,\bar{d},d]}e^{i\sum_{n,\mathbf{s}\geq 0}\left(\bar{k}_{n\mathbf{s}}d_{n\mathbf{s}0}+k_{n\mathbf{s}}\bar{d}_{n\mathbf{s}0}\right)}$$
$$=\exp\left[\sum_{n\mathbf{k}\geq 0}\begin{array}{l}\left(ik_{n\mathbf{s}}\delta_{\kappa 0}-\dfrac{g^*_{\mathbf{k}}e^{+i\mathbf{s}\cdot\mathbf{r}_0}}{\sqrt{\beta/2}}\rho_{-n-\mathbf{k}}\right)\times\dfrac{1}{-i\omega_n+\hbar\Omega_{\mathbf{k}}}\\ \times\left(i\bar{k}_{n\mathbf{s}}\delta_{\kappa 0}-\dfrac{g_{\mathbf{k}}e^{-i\mathbf{s}\cdot\mathbf{r}_0}}{\sqrt{\beta/2}}\rho_{n\mathbf{k}}\right)\end{array}\right]$$

Now substituting the above equation back to Eq. (X.1) and integrating over the $[\bar{k},k]$ field, after a few algebraic steps, we finally obtain the effective electron action as

$$S_{eff}[\bar{\psi},\psi]=S_e[\bar{\psi},\psi]-\sum_{n\mathbf{k}|\kappa\neq 0}\frac{g^*_{\mathbf{k}}g_{\mathbf{k}}\hbar\Omega_{\mathbf{k}}}{\beta(\omega_n^2+\hbar^2\Omega_{\mathbf{k}}^2)}\rho_{-n-\mathbf{k}}\rho_{n\mathbf{k}}+\sum_{\mathbf{k}n\sigma}\sum_{m\mathbf{s}}\frac{C_{\mathbf{s}}}{\sqrt{2}}\left(g_{\mathbf{s}}e^{-i\mathbf{s}\cdot\mathbf{r}_0}\bar{\psi}_{n+m\mathbf{k}+\mathbf{s}\sigma}+g^*_{\mathbf{s}}e^{i\mathbf{s}\cdot\mathbf{r}_0}\bar{\psi}_{n-m\mathbf{k}-\mathbf{s}\sigma}\right)\psi_{n\mathbf{k}\sigma} \qquad \text{(X.2)}$$

where we have used the identity $\sum_{n,\mathbf{s}\geq 0}...=\sum_{n,\mathbf{k}\geq 0}...\delta_{\kappa 0}$, and neglected a constant $\beta C_{\mathbf{s}}^2\left(-i\omega_n+\hbar\Omega_{\mathbf{s}}\right)$ term since it doesn't interact with any field. Moreover, since there are only electronic degrees of freedom left, defined in both $\mathbf{k}\geq 0$ and $\mathbf{k}\leq 0$ regions, after integrating over the dislon degrees of freedom, we don't need to restrict ourselves to the $\mathbf{k}\geq 0$ region but expect $\sum_{\mathbf{k}\geq 0}...$ can be replaced by $\frac{1}{2}\sum_{\mathbf{k}}...$ in all cases. Physically, for an electron with $\mathbf{k}\geq 0$ scattering with a dislocation, there is always a "mirrored" electron with $\mathbf{k}\leq 0$ and scattering with the dislocation with the same amplitude. Mathematically, assuming that we quantize the system with the assumption $\mathbf{k}\leq 0$ from the beginning, then, after eliminating the dislon degrees of freedom, we will obtain a system with identical physics but reside in the $\mathbf{k}\leq 0$ region. To simplify the effective Hamiltonian we have used $\sum_{\mathbf{k}\geq 0}...-\sum_{\mathbf{s}\geq 0}...=\sum_{\mathbf{k}\geq 0}...-\sum_{\mathbf{k}\geq 0}\delta_{\kappa 0}...=\frac{1}{2}\sum_{\mathbf{k}|\kappa\neq 0}...$.

Rewriting electron action Eq. (X.2) back to Hamiltonian form by normal ordering, we have the effective electron Hamiltonian

$$H_{eff}=\sum_{\mathbf{k}\sigma}(\varepsilon_{\mathbf{k}}-\mu)c^+_{\mathbf{k}\sigma}c_{\mathbf{k}\sigma}+\underbrace{\sum_{\mathbf{kpq}\sigma\sigma'}V_{eff}(\mathbf{k})c^+_{\mathbf{k}+\mathbf{p}\sigma}c^+_{\mathbf{k}-\mathbf{q}\sigma'}c_{\mathbf{q}\sigma'}c_{\mathbf{p}\sigma}}_{\text{quantum}}+\underbrace{\frac{1}{L^2}\sum_{\mathbf{k}\sigma}\sum_{\mathbf{s}}\left(A_{\mathbf{s}}e^{-i\mathbf{s}\cdot\mathbf{r}_0}c^+_{\mathbf{k}+\mathbf{s}\sigma}+A^*_{\mathbf{s}}e^{i\mathbf{s}\cdot\mathbf{r}_0}c^+_{\mathbf{k}-\mathbf{s}\sigma}\right)c_{\mathbf{k}\sigma}}_{\text{classical}} \qquad \text{(X.3)}$$

where the coupling coefficients are written as

$$V_{eff}(\mathbf{k})\sim-\frac{g^*_{\mathbf{k}}g_{\mathbf{k}}}{\hbar\Omega_{\mathbf{k}}}=-\left(\frac{N}{L^5}\right)^2\frac{(eV_{\mathbf{k}})^2}{2m_{\mathbf{k}}\Omega_{\mathbf{k}}^2}\left[\mathbf{k}\cdot\mathbf{F}(\mathbf{k})\right]^2$$

and $A_{\mathbf{s}}\equiv\frac{g_{\mathbf{s}}C_{\mathbf{s}}}{\sqrt{2}}=\frac{eN}{2L^3}V_{\mathbf{s}}\left[i\mathbf{s}\cdot\mathbf{F}(\mathbf{s})\right]$.

To recap a bit, in these expressions of $V_{eff}(\mathbf{k})$ and $A_{\mathbf{s}}$, *N* is the total number of atoms, *L* is the system size, $m_{\mathbf{k}}\equiv T_{\mathbf{k}}/L$, $\Omega_{\mathbf{k}}=\sqrt{W_{\mathbf{k}}/T_{\mathbf{k}}}$ is the dislon dispersion, with $T_{\mathbf{k}}$ and $W_{\mathbf{k}}$ defined in Eq. (IV.1), $\mathbf{F}(\mathbf{k})$ is defined in Eq. (III.11), $V_{\mathbf{k}}$ is the Coulomb interaction defined after Eq. (VI.1), and $\mathbf{s}$ is the 2D version of $\mathbf{k}$ perpendicular to the dislocation line direction.

Eq. (X.3) has clear physical meaning, indicating that the electron-dislocation interaction has two distinct effects: the *classical* effect is the quenched dislocation scattering, which can change the electron momentum perpendicular to the dislocation line direction (noticing **s** is the 2D wavevector perpendicular to the dislocation line direction), while the *quantum* effect can mediate a coupling between two electrons forming a Cooper pair, just as a phonon does. As one early application of the dislon theory, it has been shown that the competition between the classical and quantum effects can determine the shift of the transition temperature in a dislocated superconductor [76].

## XI. EFFECTIVE PHONON THEORY WITH FLUTTERING INTERACTION

We define the effective action of phonon as

$$e^{-S_{eff}[\bar{b},b]} \equiv e^{-S_{ph}[\bar{b},b]} \times \int D[\bar{f},f] e^{-S_{dis}[\bar{f},f]-S_{flu}[\bar{b},b,\bar{f},f]} \quad \text{(XI.1)}$$

in which $S_{dis}[\bar{f},f] = \sum_{n,\mathbf{k}\geq 0} \bar{f}_{\mathbf{k}n}(-i\omega_n + \hbar\Omega_{\mathbf{k}}) f_{\mathbf{k}n}$ is the *f*-field of the dislon, coupling with a phonon through the fluttering mechanism.

Now integrating over the $[\bar{f},f]$ field, we have

$$\int D[\bar{f},f] e^{-S_{dis}[\bar{f},f]-S_{flu}[\bar{b},b,\bar{f},f]} = \exp\left( \sum_{\mathbf{k}\geq 0,n} \frac{\hbar^2 \rho\omega_{\mathbf{k}}\Omega_{\mathbf{k}}}{2T_{\mathbf{k}}} [\boldsymbol{\varepsilon}_{\mathbf{k}} \cdot \mathbf{F}(\mathbf{k})]^2 \times \left(-b_{\mathbf{k}n} + \bar{b}_{-\mathbf{k},-n}\right) \frac{1}{-i\omega_n + \hbar\Omega_{\mathbf{k}}} \left(b_{-\mathbf{k},-n} - \bar{b}_{\mathbf{k}n}\right) \right)$$

from which the effective phonon action Eq. (XI.1) can be written directly as

$$S_{eff}[\bar{b},b] \equiv S_{ph}[\bar{b},b] - \sum_{\mathbf{k}\geq 0,n} \frac{\rho\hbar\omega_{\mathbf{k}}}{4T_{\mathbf{k}}} \frac{[\boldsymbol{\varepsilon}_{\mathbf{k}} \cdot \mathbf{F}(\mathbf{k})]^2 \hbar^2\Omega_{\mathbf{k}}^2}{\omega_n^2 + \hbar^2\Omega_{\mathbf{k}}^2} \times \left(b_{\mathbf{k}n} - \bar{b}_{-\mathbf{k},-n}\right)\left(-b_{-\mathbf{k},-n} + \bar{b}_{\mathbf{k}n}\right) \quad \text{(XI.2)}$$

where we have used $\sum_{\mathbf{k}\geq 0} ... = \frac{1}{2}\sum_{\mathbf{k}} ...$ after eliminating the dislon degrees of freedom. Now performing Keldysh rotation, defining that

$$\begin{aligned} \phi_{\mathbf{k}n} &= \frac{1}{\sqrt{2}}\left(b_{\mathbf{k}n} + \bar{b}_{-\mathbf{k}-n}\right), \ \bar{\phi}_{\mathbf{k}n} = \frac{1}{\sqrt{2}}\left(\bar{b}_{\mathbf{k}n} + b_{-\mathbf{k}-n}\right) \\ \pi_{\mathbf{k}n} &= \frac{1}{\sqrt{2}}\left(b_{\mathbf{k}n} - \bar{b}_{-\mathbf{k}-n}\right), \ \bar{\pi}_{\mathbf{k}n} = \frac{1}{\sqrt{2}}\left(\bar{b}_{\mathbf{k}n} - b_{-\mathbf{k}-n}\right) \end{aligned} \quad \text{(XI.3)}$$

then using the equality $\bar{\phi}_{\mathbf{k}n}\phi_{\mathbf{k}n} + \bar{\phi}_{\mathbf{k}n}\pi_{\mathbf{k}n} + \bar{\pi}_{\mathbf{k}n}\phi_{\mathbf{k}n} + \bar{\pi}_{\mathbf{k}n}\pi_{\mathbf{k}n} = 2\bar{b}_{\mathbf{k}n}b_{\mathbf{k}n}$, the free-phonon action Eq. (IX.6) can be rewritten as

$$\begin{aligned} S_{ph}[\bar{\phi},\phi,\bar{\pi},\pi] &= \sum_{n\mathbf{k}} \bar{b}_{\mathbf{k}n}(-i\omega_n + \hbar\omega_{\mathbf{k}}) b_{\mathbf{k}n} \\ &= \frac{1}{2}\sum_{n\mathbf{k}} \left(-i\omega_n + \hbar\omega_{\mathbf{k}}\right) \begin{pmatrix} \bar{\phi}_{\mathbf{k}n} & \bar{\pi}_{\mathbf{k}n} \end{pmatrix} \begin{pmatrix} 1 & 1 \\ 1 & 1 \end{pmatrix} \begin{pmatrix} \phi_{\mathbf{k}n} \\ \pi_{\mathbf{k}n} \end{pmatrix} \end{aligned} \quad \text{(XI.4)}$$

Now, noticing that $\bar{\phi}_{\mathbf{k}n} = \phi_{-\mathbf{k}-n}$, the fields $\phi$ and $\pi$ are indeed real scalar fields, i.e. the complex scalar phonon fields $\bar{b}, b$ are decomposed into a displacement-like field $\phi$ and a momentum-like field $\pi$.

Now the effective phonon action can be written as

$$S_{eff}[\bar{\phi},\phi,\bar{\pi},\pi] = \sum_{\mathbf{k}n} \begin{pmatrix} \bar{\phi}_{\mathbf{k}n} & \bar{\pi}_{\mathbf{k}n} \end{pmatrix} \times \left[ \frac{1}{2} D_{0n\mathbf{k}}^{-1} \begin{pmatrix} 1 & 1 \\ 1 & 1 \end{pmatrix} - J_{n\mathbf{k}} \begin{pmatrix} 0 & 0 \\ 0 & 1 \end{pmatrix} \right] \begin{pmatrix} \phi_{\mathbf{k}n} \\ \pi_{\mathbf{k}n} \end{pmatrix} \quad \text{(XI.5)}$$

where the coupling constant of phonon-dislon fluttering is defined as $J_{n\mathbf{k}} \equiv \frac{\rho\hbar\omega_{\mathbf{k}}}{2T_{\mathbf{k}}} \frac{[\boldsymbol{\varepsilon}_{\mathbf{k}} \cdot \mathbf{F}(\mathbf{k})]^2 \hbar^2\Omega_{\mathbf{k}}^2}{\omega_n^2 + \hbar^2\Omega_{\mathbf{k}}^2}$, and the free-phonon propagator $D_{0n\mathbf{k}}$ is defined as $D_{0n\mathbf{k}} \equiv \frac{1}{-i\omega_n + \hbar\omega_{\mathbf{k}}}$.

Diagonalizing the 2x2 matrix in Eq. (XI.5), we have

$$S_{eff}[\bar{\phi}',\phi',\bar{\pi}',\pi'] = \frac{1}{2}\sum_{\mathbf{k}n} \begin{pmatrix} \bar{\phi}'_{\mathbf{k}n} & \bar{\pi}'_{\mathbf{k}n} \end{pmatrix} \begin{pmatrix} D_{n\mathbf{k}+}^{-1} & 0 \\ 0 & D_{n\mathbf{k}-}^{-1} \end{pmatrix} \begin{pmatrix} \phi'_{\mathbf{k}n} \\ \pi'_{\mathbf{k}n} \end{pmatrix} \quad \text{(XI.6)}$$

where we use the notation $[\bar{\phi}',\phi',\bar{\pi}',\pi']$ to denote the diagonalized fields, and $D_{n\mathbf{k}\pm}^{-1} \equiv D_{0n\mathbf{k}}^{-1} - J_{n\mathbf{k}} \pm \sqrt{D_{0n\mathbf{k}}^{-2} + J_{n\mathbf{k}}^2}$. Without dislon-phonon coupling, i.e. $J_{n\mathbf{k}} = 0$, Eq. (XI.6) should reduce back to the free phonon propagator $D_{0n\mathbf{k}}^{-1}$ directly instead of giving a null result, which enables us to safely neglect the unphysical $D_{n\mathbf{k}-}^{-1}$; finally, we obtain the diagonalized effective phonon theory in 3D

$$S_{eff}[\bar{\phi}',\phi'] = \frac{1}{2}\sum_{\mathbf{k}n} \bar{\phi}'_{\mathbf{k}n} \left[ D_{0n\mathbf{k}}^{-1} - J_{n\mathbf{k}} + \sqrt{D_{0n\mathbf{k}}^{-2} + J_{n\mathbf{k}}^2} \right] \phi'_{\mathbf{k}n} \quad \text{(XI.7)}$$

Notice that here, $\phi'_{\mathbf{k}n}$ after rotation does not resemble a displacement operator, but is more like an annihilation operator.

With this effective theory, a number of problems associated with dislocation-phonon interaction can be solved directly, including how dislocations may change phonon dispersion, and how the heat capacity in a dislocated crystal is changed, etc.

## XII. EFFECTIVE PHONON THEORY WITH ANHARMONICITY

To obtain an effective phonon theory with anharmonicity, we first define a composite phonon displacement operator $\Phi_{\mathbf{k}n} \equiv \sum_{\mathbf{k}_1\mathbf{k}_2,p,q} A(\mathbf{k}_1,\mathbf{k}_2,\mathbf{k}) \delta_{p+q+n,0} \phi_{\mathbf{k}_1,p}\phi_{\mathbf{k}_2,q}$; then Eq. (IX.8) can be rewritten in terms of a simpler form as

$$S_{Anh}[\bar{\phi},\phi,\bar{d},d] = \frac{\hbar}{2\rho\sqrt{\beta}L^2} \sum_{\mathbf{k}>0,n} \left[ \Phi_{\mathbf{k}n} d_{\mathbf{k}n} + \Phi_{-\mathbf{k}-n} \bar{d}_{\mathbf{k}n} \right] \quad \text{(XII.1)}$$

One thing worth mentioning is that

1) In the fluttering interaction, it is the dislon *f*-field which couples with the phonon $\pi$ field.
2) In the anharmonic interaction, it is the dislon *d*-field which couples with the phonon $\phi$ field.

This fact can be used to greatly simplify the phonon effective theory, as the fluttering and anharmonicity can be treated independently.

Assume that we only plan to study a complete phonon-dislon interaction system without considering electronic degrees of freedom, and define the effective action of phonon coupling with the *d*-field, $S_{eff}[\bar{\phi},\phi]$, as

$$e^{-S_{eff}[\bar{\phi},\phi]} \equiv e^{-S_{ph}[\bar{\phi},\phi]}\int D[\bar{d},d]e^{-S_{dis}[\bar{d},d]-S_{anh}[\bar{\phi},\phi,\bar{d},d]} \times \prod_{n\mathbf{s}\geq 0}\delta(d_{n\mathbf{s}0}-\sqrt{\beta}C_{\mathbf{s}})\delta(\bar{d}_{n\mathbf{s}0}-\sqrt{\beta}C_{\mathbf{s}}) \qquad \text{(XII.2)}$$

Using the same method implemented in Section X, performing a Fourier transform of the δ-function and integrating over the dislon degrees of freedom so that $\delta(d_{n\mathbf{s}0}-\sqrt{\beta}C_{\mathbf{s}})=\int\frac{d\bar{k}_{n\mathbf{s}}}{2\pi}e^{i\bar{k}_{n\mathbf{s}}(d_{n\mathbf{s}0}-\sqrt{\beta}C_{\mathbf{s}})}$ , and defining the functional measure $D[\bar{k},k]\equiv\prod_{n\mathbf{s}}\int\frac{d\bar{k}_{n\mathbf{s}}}{2\pi}\frac{dk_{n\mathbf{s}}}{2\pi}$, we have

$$e^{-S_{eff}[\bar{\phi},\phi]} = e^{-S_{ph}[\bar{\phi},\phi]}\int D[\bar{d},d]D[\bar{k},k]e^{-i\sqrt{\beta}C_{\mathbf{s}}\sum_{\mathbf{s}\geq 0,n}\left(k_{n\mathbf{s}}+\bar{k}_{n\mathbf{s}}\right)} \times e^{i\sum_{\mathbf{k}\geq 0,n}\left(\bar{k}_{n\mathbf{s}}\delta_{\kappa 0}d_{n\mathbf{k}}+k_{n\mathbf{s}}\delta_{\kappa 0}\bar{d}_{n\mathbf{k}}\right)} e^{-S_{dis}[\bar{d},d]-S_{anh}[\bar{\phi},\phi,\bar{d},d]}$$

Now integrating over the dislon degrees of freedom, we have

$$\int D[\bar{d},d]e^{i\sum_{\mathbf{k}\geq 0,n}\left(\bar{k}_{n\mathbf{s}}\delta_{\kappa 0}d_{n\mathbf{k}}+k_{n\mathbf{s}}\delta_{\kappa 0}\bar{d}_{n\mathbf{k}}\right)} e^{-S_{dis}[\bar{d},d]-S_{anh}[\bar{\phi},\phi,\bar{d},d]}$$
$$=\exp\left(\sum_{\mathbf{k}\geq 0,n}\begin{pmatrix}\left(i\bar{k}_{n\mathbf{s}}\delta_{\kappa 0}-\frac{\hbar}{2\rho\sqrt{\beta}L^2}\Phi_{\mathbf{k}n}\right)\times \\ \frac{1}{-i\omega_n+\hbar\Omega_{\mathbf{k}}}\times\left(ik_{n\mathbf{s}}\delta_{\kappa 0}-\frac{\hbar}{2\rho\sqrt{\beta}L^2}\Phi_{-\mathbf{k}-n}\right)\end{pmatrix}\right)$$

Now further defining the 2D version of the composite operator $\Phi_{\mathbf{s},n}\equiv\sum_{\mathbf{k}_1\mathbf{k}_2,p,q}A\left(\mathbf{k}_1,\mathbf{k}_2,(\mathbf{s},\kappa=0)\right)\delta_{p+q,-n}\phi_{\mathbf{k}_1,p}\phi_{\mathbf{k}_2,q}$ and integrating over the $[\bar{k},k]$ fields, after a few steps, we obtain

$$S_{eff}[\bar{\phi},\phi]\equiv S_{ph}[\bar{\phi},\phi]+\frac{\hbar}{4\rho L^2}\sum_{\mathbf{s},n}\left(C_{\mathbf{s}}\Phi_{\mathbf{s},n}+C_{\mathbf{s}}\Phi_{-\mathbf{s},-n}\right) -\frac{\hbar^2}{8\beta\rho^2L^4}\sum_{\mathbf{k},n}\frac{\hbar\Omega_{\mathbf{k}}}{\omega_n^2+\hbar^2\Omega_{\mathbf{k}}^2}\Phi_{\mathbf{k}n}\Phi_{-\mathbf{k}-n} \qquad \text{(XII.3)}$$

where we have seen that both the 2nd order quadratic term of displacement and 4th order quartic term exist.

At this step, Eq. (XII.3) becomes a quadratic theory of the composite operator $\Phi_{\mathbf{k}n}$ . To link back to phonon displacement properties, we first rewrite Eq. (XII.3) in terms of phonon displacement fields as

$$S_{eff}[\bar{\phi},\phi]\equiv S_{ph}[\bar{\phi},\phi] +\frac{\hbar}{4\rho L^2}\sum_{\mathbf{s},n}\begin{pmatrix}C_{\mathbf{s}}\sum_{\mathbf{k}_1\mathbf{k}_2,pq}A_{\mathbf{k}_1,\mathbf{k}_2,\mathbf{s}}\delta_{p+q+n,0}\phi_{\mathbf{k}_1,p}\phi_{\mathbf{k}_2,q}+ \\ C_{\mathbf{s}}\sum_{\mathbf{k}_1\mathbf{k}_2,pq}A_{\mathbf{k}_1,\mathbf{k}_2,-\mathbf{s}}\delta_{p+q-n,0}\phi_{\mathbf{k}_1,p}\phi_{\mathbf{k}_2,q}\end{pmatrix} -\frac{\hbar^2}{8\beta\rho^2L^4}\sum_{\mathbf{k},n}\frac{\hbar\Omega_{\mathbf{k}}}{\omega_n^2+\hbar^2\Omega_{\mathbf{k}}^2}\times \sum_{\substack{\mathbf{k}_1\mathbf{k}_2,pq\\ \mathbf{k}'_1\mathbf{k}'_2,p'q'}}A_{\mathbf{k}_1,\mathbf{k}_2,-\mathbf{k}}\delta_{p+q,n}A_{\mathbf{k}'_1,\mathbf{k}'_2,\mathbf{k}}\delta_{p'+q',-n}\phi_{\mathbf{k}_1,p}\phi_{\mathbf{k}_2,q}\phi_{\mathbf{k}'_1,p'}\phi_{\mathbf{k}} \qquad \text{(XII.4)}$$

Now we examine the coefficients in the quadratic term. If we only look at $C_{\mathbf{s}}$ , since $\lim_{\mathbf{s}\to 0}C_{\mathbf{s}}=0$, this makes the diagonal component ( $\mathbf{k}_1=-\mathbf{k}_2$ ) of the quadratic scattering term *seemingly* vanish. However, after a closer look, the combined coefficient $C_{\mathbf{s}}A(\mathbf{k}_1,\mathbf{k}_2,\mathbf{s})$ can be simplified using Eqs. (V.7) and (VIII.5) as

$$C_{\mathbf{s}}A(\mathbf{k}_1,\mathbf{k}_2,\mathbf{s})=\delta^{(3)}(\mathbf{k}_1+\mathbf{k}_2+\mathbf{s})V^{abc}_{\mathbf{k}_1\mathbf{k}_2\mathbf{s}} \times\boldsymbol{\varepsilon}^a_{\mathbf{k}_1}\boldsymbol{\varepsilon}^b_{\mathbf{k}_2}\mathbf{F}_c(\mathbf{s})\sqrt{\frac{1}{\omega_{\mathbf{k}_1}\omega_{\mathbf{k}_2}}} \qquad \text{(XII.5)}$$

Which gives $\lim_{\mathbf{s}\to 0}C_{\mathbf{s}}A(\mathbf{k}_1,\mathbf{k}_2,\mathbf{s})\to\infty$ since $\mathbf{F}(\mathbf{s})\sim\pm\frac{b}{s^2}$ . Therefore, the quadratic scattering terms (2nd line in Eq. (XII.4) ) is maximized in the $\mathbf{s}\to 0$ limit, instead of approaching 0. Therefore, we can only keep the dominating diagonal term with $\mathbf{s}$=0 , and thus $-\mathbf{k}_1=\mathbf{k}_2\equiv\mathbf{k}$ . Since in reality, $\mathbf{F}(\mathbf{s}=0)$ should not be divergent but contain a finite infrared long-wavelength physical cutoff, we further define a coefficient $\Delta_{\mathbf{k}}$ as $\Delta_{\mathbf{k}}\equiv -V^{abc}_{-\mathbf{k},\mathbf{k}0}\boldsymbol{\varepsilon}^a_{-\mathbf{k}}\boldsymbol{\varepsilon}^b_{\mathbf{k}}\mathbf{F}_c(0)$ , where $\mathbf{F}_c(0)\equiv\mathbf{F}_c\left(\mathbf{s}=0\right)$.

At this stage, the theoretical construction of dislon theory interacting with electrons and phonons is complete. With the new effective electron Hamiltonian Eq. (X.3), the influence of dislocations on materials' electronic and optical properties can be studied; with the effective phonon action Eq. (XII.4), the influence of dislocations on materials' phononic and thermal properties can be studied. A new Hamiltonian or new actions means new opportunity. In this sense, we do not intend to exhaust all these opportunities; instead, we demonstrate the power of the dislon theory by providing a few case studies in the next few sections: computing the electron relaxation time for scattering with dislocations, the electrical conductivity caused by electron-dislocation scattering, and the thermal conductivity arising from phonon-dislocation interaction.

## XIII. THE RELAXATION TIME FOR ELECTRON-DISLON SCATTERING

Since the dislon $d$-field has a boundary condition Eq.(V.7), to compute transport properties, instead of directly applying the Feynman diagrams of electron-dislon interaction, which is done in the case of studying electron-phonon coupling, the effective electron theory Eq. (X.3) should be used in order to take into account the boundary condition properly. In this section, we focus on the classical scattering between an electron and dislon from Eq. (X.3), i.e.

$$H_{eff} = \sum_{\mathbf{k}\sigma}(\varepsilon_{\mathbf{k}} - \mu) c^{+}_{\mathbf{k}\sigma} c_{\mathbf{k}\sigma} + \frac{1}{L^2}\sum_{\mathbf{k}\sigma}\sum_{\mathbf{s}} A_{\mathbf{s}} e^{-i\mathbf{s}\cdot\mathbf{r}_0} c^{+}_{\mathbf{k}+\mathbf{s}\sigma} c_{\mathbf{k}\sigma} \quad \text{(XIII.1)}$$

where we have used $A^*_{\mathbf{s}} = A_{-\mathbf{s}}$ to simplify the quadratic part of Eq. (X.3), and written the scattering amplitude as $A_{\mathbf{s}} = \frac{g_{\mathbf{s}} C_{\mathbf{s}}}{\sqrt{2}} = \frac{eN}{2L^3} V_{\mathbf{s}}\left[i\mathbf{s}\cdot\mathbf{F}(\mathbf{s})\right]$ (distinguish this 1-index scattering amplitude from the anharmonic coupling coefficient $A(\mathbf{k}_1, \mathbf{k}_2, \mathbf{k}_3)$ which has 3 indices). It is also worth mentioning that the Fourier transform of $A_{\mathbf{s}}$ gives the scattering potential of classical deformation potential scattering, shown in an early study of dislon theory [76].

In the following part of this section, we generalize Eq. (XIII.1) to multiple dislocation lines, using a similar technique of studying impurity scattering [96], since both impurity scattering and electron- classical dislocation scattering are quenched, quadratic potential scattering.

For parallel dislocation arrays with dislocation core position $\mathbf{r}_j \equiv (x_j, y_j)$, $j = 1, 2, ..., N_{dis}$, the electron effective Hamiltonian $H_{eff}$ taking into account the parallel dislocations scattering can be written as

$$\begin{aligned} H_{eff} &= H_0 + H_{int} \\ &= \sum_{\mathbf{k}\sigma}(E_{\mathbf{k}} - \mu) c^{+}_{\mathbf{k}\sigma} c_{\mathbf{k}\sigma} + \frac{1}{L^2}\sum_{j=1}^{N_{dis}}\sum_{\mathbf{k}\sigma\mathbf{s}} A_{\mathbf{s}} e^{-i\mathbf{s}\cdot\mathbf{r}_j} c^{+}_{\mathbf{k}+\mathbf{s}\sigma} c_{\mathbf{k}\sigma} \end{aligned} \quad \text{(XIII.2)}$$

Written in this form, we can utilize the standard quantum many-body approach [81] to study the electron-dislocation array scattering problem. Define the imaginary time Green's function as $G(\mathbf{k}\tau; \mathbf{k}'\tau') \equiv -\left\langle T_\tau c_{\mathbf{k}}(\tau) c^{+}_{\mathbf{k}'}(\tau')\right\rangle$, and non-interacting Green's function as $G_0(\mathbf{k}, \tau - \tau') = -\left\langle T_\tau c_{\mathbf{k}}(\tau) c^{+}_{\mathbf{k}}(\tau')\right\rangle_0$ (diagonal) associated with the non-interacting Hamiltonian $H_0$, in which $T_\tau$ is the time-ordering operator. We further define the corresponding Fourier transformed Green's functions in the Matsubara frequency domain as

$$\begin{aligned} G(\mathbf{k}, \mathbf{k}'; p_n) &= \int_0^\beta G(\mathbf{k}\tau; \mathbf{k}'\tau') e^{ip_n(\tau - \tau')} d\tau \\ G(\mathbf{k}\tau; \mathbf{k}'\tau') &= \frac{1}{\beta}\sum_n G(\mathbf{k}, \mathbf{k}'; p_n) e^{-ip_n(\tau - \tau')} \end{aligned} \quad \text{(XIII.3)}$$

Then, we have the Dyson's equation (Appendix B)

$$\begin{aligned} G(\mathbf{k}, \mathbf{k}'; p_n) &= G_0(\mathbf{k}, p_n) + \frac{1}{L^2} \times \\ &\sum_{j=1}^{N_{dis}}\sum_{\mathbf{q}} e^{-i(\mathbf{k}-\mathbf{q})\cdot\mathbf{r}_j} G_0(\mathbf{k}, p_n) A_{\mathbf{k}-\mathbf{q}} \delta_{k_z, q_z} G(\mathbf{q}, \mathbf{k}'; p_n) \end{aligned} \quad \text{(XIII.4)}$$

Here we need to bear in mind that the "**s**" is a 2D vector perpendicular to the dislocation direction, instead of a 3D vector, resulting in the $\delta_{k_z, q_z}$ coefficient. In other words, classical electron-dislocation scattering will not change the electron momentum along the dislocation line direction, which is quite reasonable.

At this stage, we use diagrammatic tools to recap the problem:

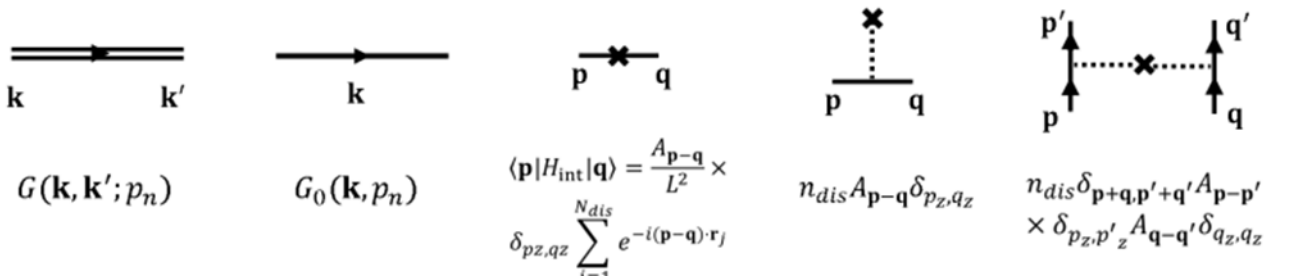

Then, Eq. (XIII.4) can be rewritten pictorially as:

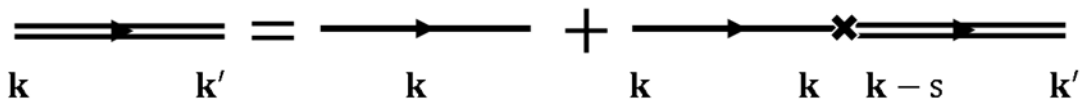

and a few low order Green's functions can be written in a diagrammatic way as

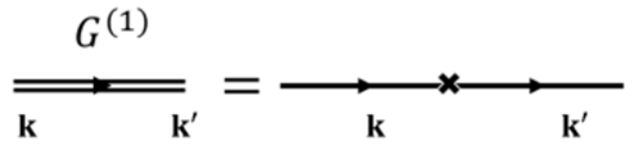

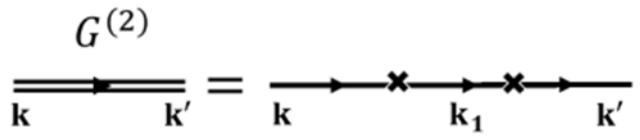

which correspond to following expressions

$$\begin{aligned} G^{(1)}(\mathbf{k}, \mathbf{k}'; p_n) &= \frac{1}{L^2}\sum_{j_1=1}^{N_{dis}} e^{-i(\mathbf{k}-\mathbf{k}')\cdot\mathbf{r}_{j_1}} G_0(\mathbf{k}, p_n) \times \\ &\quad A_{\mathbf{k}-\mathbf{k}'} \delta_{k_z, k'_z} G_0(\mathbf{k}', p_n) \\ G^{(2)}(\mathbf{k}, \mathbf{k}'; p_n) &= \left(\frac{1}{L^2}\right)^2 \sum_{\substack{j_1=1 \\ j_2=1}}^{N_{dis}}\sum_{\mathbf{k}_1} e^{-i(\mathbf{k}-\mathbf{k}_1)\cdot\mathbf{r}_{j_1}} e^{-i(\mathbf{k}_1-\mathbf{k}')\cdot\mathbf{r}_{j_2}} \\ &\times G_0(\mathbf{k}, p_n) A_{\mathbf{k}-\mathbf{k}_1} \delta_{k_z, k_{1z}} G_0(\mathbf{k}_1, p_n) A_{\mathbf{k}_1-\mathbf{k}'} \delta_{k'_z, k_{1z}} G_0(\mathbf{k}', p_n) \end{aligned} \quad \text{(XIII.5)}$$

from which we could deduce the corresponding Feynman rules for electron–classical dislocation scattering, that

a) for each scattering vertex, 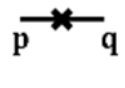 is denoted by

$$\langle \mathbf{p}| H_{\text{int}} |\mathbf{q}\rangle = A_{\mathbf{p}-\mathbf{q}} \delta_{pz,qz} \frac{1}{L^2}\sum_{j=1}^{N_{dis}} e^{-i(\mathbf{p}-\mathbf{q})\cdot\mathbf{r}_j}$$

b) summation over all internal momenta is performed as usual.

Up to this step, no averaging process hastaken place yet. However, when the dislocation core locations are purely random, in the regime that the sample size $d$ is much greater than the electron phase coherence length $l_\phi$, i.e. $d \gg l_\phi$, we can further consider the dislocation distribution as homogeneous and hence average over all possible configurations, i.e.

$$\left\langle \sum_{j=1}^{N_{dis}} e^{-i(\mathbf{k}-\mathbf{k}')\cdot\mathbf{r}_j} \right\rangle = N_{dis}\delta_{\mathbf{k}\mathbf{k}'} \qquad \text{(XIII.6)}$$

After this impurity averaging process, Eq. (XIII.5) can be rewritten as

$$\begin{aligned}
&\left\langle G^{(1)}(\mathbf{k},\mathbf{k}';p_n)\right\rangle = n_{dis}\delta_{\mathbf{k},\mathbf{k}'}A_{\mathbf{k}=0}G_0^2(\mathbf{k},p_n)\\
&\left\langle G^{(2)}(\mathbf{k},\mathbf{k}';p_n)\right\rangle^{j_1\neq j_2} = n_{dis}^2\delta_{\mathbf{k},\mathbf{k}'}A_{\mathbf{k}=0}^2G_0^3(\mathbf{k},p_n)\\
&\left\langle G^{(2)}(\mathbf{k},\mathbf{k}';p_n)\right\rangle^{j_1=j_2} = n_{dis}\delta_{\mathbf{k},\mathbf{k}'}G_0^2(\mathbf{k},p_n)\times\\
&\frac{1}{L^2}\sum_{\mathbf{k}_1}\left|A_{\mathbf{k}-\mathbf{k}_1}\right|^2\delta_{k_z,k_{1z}}G_0(\mathbf{k}_1,p_n)
\end{aligned} \qquad \text{(XIII.7)}$$

The corresponding Feynman rules taking, into account the dislocation averaging, can be written as

a) for each vertex ✖, we have $n_{dis}\delta_{\sum\mathbf{p}_{in},\sum\mathbf{p}_{out}}$, i.e. the dislocation density $n_{dis}$ starts to appear.

b) for each interaction line 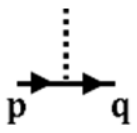, we have $A_{\mathbf{p}-\mathbf{q}}\delta_{pz,qz}$.

The expressions in Eq. (XIII.7) can be rewritten in a pictorial way as

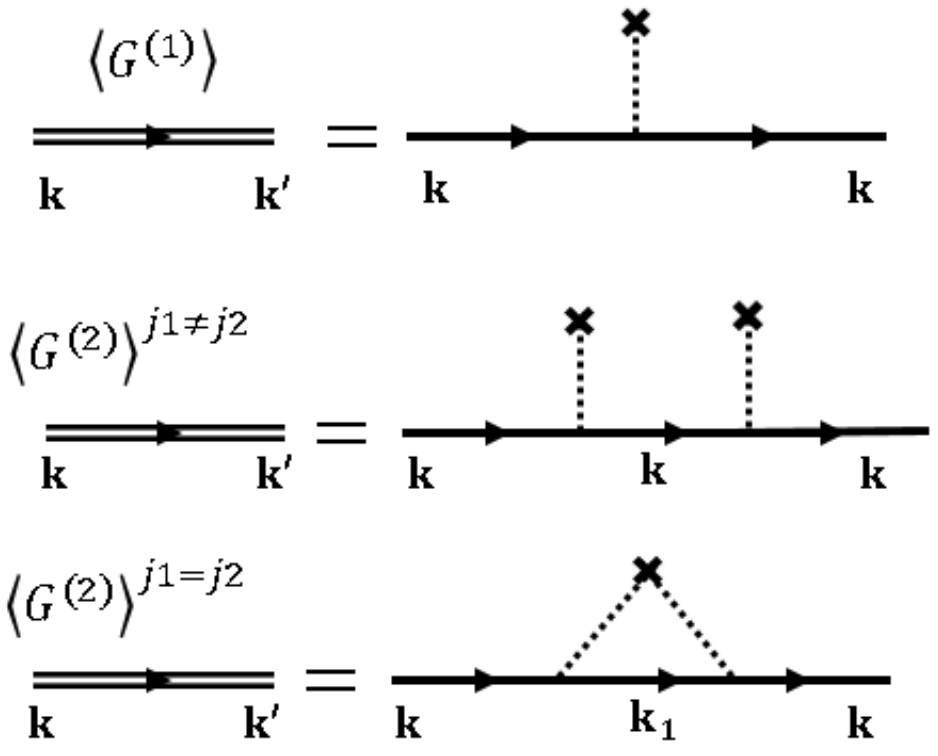

and we can build an arbitrarily complicated diagram, for example the following diagram

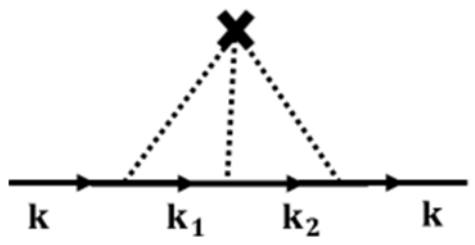

$$= n_{dis}G_0^2(\mathbf{k},p_n)\frac{1}{L^4}\sum_{\mathbf{k}_1\mathbf{k}_2}\begin{matrix}A_{\mathbf{k}-\mathbf{k}_1}\delta_{k_z,k_{1z}}G_0(\mathbf{k}_1,p_n)\times\\ A_{\mathbf{k}_1-\mathbf{k}_2}\delta_{k_{1z},k_{2z}}G_0(\mathbf{k}_2,p_n)A_{\mathbf{k}_2-\mathbf{k}}\delta_{k_{2z},k_z}\end{matrix}$$

At this stage, the dislocation averaging scheme is complete, from which we can compute the self-energy and relaxation time for electron classical dislocation scattering to arbitrarily high order.

Self-energy $\Sigma$ is composed of one-particle irreducible (1PI) diagrams with external legs amputated, and we have the following equation [81]

$$\left\langle G(\mathbf{k};ip_n)\right\rangle = \frac{1}{ip_n - E_{\mathbf{k}} + \mu - \Sigma(\mathbf{k},ip_n)} \qquad \text{(XIII.8)}$$

valid generally. With 1st-order Born approximation (FOBA), we write the following self-energy diagram

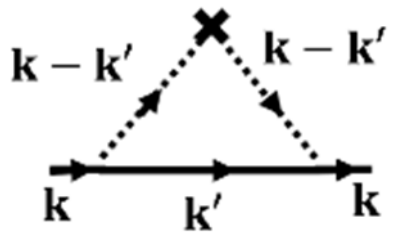

$$= \Sigma^{FOBA}(\mathbf{k},ip_n) = n_{dis}\frac{1}{L^2}\sum_{\mathbf{k}'}\left|A_{\mathbf{k}-\mathbf{k}'}\right|^2\delta_{kz,k'z}G_0(\mathbf{k}';ip_n)$$

$$= n_{dis}\frac{1}{L^2}\sum_{\mathbf{k}'}\frac{\left|A_{\mathbf{k}-\mathbf{k}'}\right|^2\delta_{kz,k'z}}{ip_n - E_{\mathbf{k}'} + \mu}$$

After analytical continuation (assuming infinitesimal $\eta$ and $p_n > 0$) from Matsubara frequency to real frequency $\omega$, we have

$$\Sigma^{FOBA}(\mathbf{k},\omega) = n_{dis}\frac{1}{L^2}\sum_{\mathbf{k}'}\frac{\left|A_{\mathbf{k}-\mathbf{k}'}\right|^2\delta_{kz,k'z}}{\omega - E_{\mathbf{k}'} + \mu + i\eta} \qquad \text{(XIII.9)}$$

Hence the relaxation time $\tau_{\mathbf{k}}$ of electron – parallel dislocation array scattering can be written as

$$\begin{aligned}
\frac{\hbar}{\tau_{\mathbf{k}}} &= 2\,\mathrm{Im}\left(\Sigma^{FOBA}\left(\mathbf{k},\omega=\varepsilon_{\mathbf{k}}-\mu\right)\right)\\
&= \frac{2\pi n_{dis}}{L^2}\sum_{\mathbf{k}'}\left|A_{\mathbf{k}-\mathbf{k}'}\right|^2\delta_{kz,k'z}\delta(\varepsilon_{\mathbf{k}}-\varepsilon_{\mathbf{k}'})
\end{aligned} \qquad \text{(XIII.10)}$$

where we have recovered the normal Fermi's Golden rule. The benefit of taking this approach is to explore the scattering schemes to arbitrarily high order, or to take into account any electron-electron correlation effects. For instance, the possibility of weak localization arising from coherent superposition between scattering paths of impurities [83] – now for dislocation arrays – can be properly addressed using this approach. In any case, we may consider Eq. (XIII.10) as a good starting point to demonstrate the power of the dislon theory to compute electronic transport properties with parallel dislocation arrays.

## XIV. ELECTRICAL CONDUCTIVITY WITH PARALLEL DISLOCATION ARRAYS

In this section, we demonstrate another usage of dislon theory to compute the electrical conductivity for electron – parallel dislocation array scattering, at a full quantum level (beyond semi-classical approach). The electrical conductivity tensor $\sigma_{\alpha\beta}(\mathbf{q},\omega)$ through the Kubo formula follows largely the standard many-body approach [81,83,96], and we apply the generic scheme of computing $\sigma_{\alpha\beta}(\mathbf{q},\omega)$ to the case of dislocation arrays:

$$\mathrm{Re}\,\sigma_{ab}(\mathbf{q},\omega)=-\frac{e^2}{\omega}\mathrm{Im}\,\Pi^R_{ab}(\mathbf{q},\omega) \quad \text{(XIV.1)}$$

where $a,b=1,2,3$ are the Cartesian coordinates, and $\Pi^R_{\alpha\beta}$ is the retarded current-current correlation function, which can be obtained by analytical continuation of the Matsubara current-current correlation function $\Pi_{ab}(\mathbf{q},i\omega_n)$ :

$$\Pi^R_{ab}(\mathbf{q},\omega)=\Pi_{ab}(\mathbf{q},i\omega_n\to\omega+i\eta) \quad \text{(XIV.2)}$$

in which $\omega_n$ is the Bosonic Matsubara frequency. Hence, the central quantity of computing conductivity is to compute the Matsubara current-current correlation function

$$\Pi_{ab}(\mathbf{q},i\omega_n)=-\frac{1}{\beta L^3}\left\langle J_a(\mathbf{q},i\omega_n)J_b(-\mathbf{q},-i\omega_n)\right\rangle \quad \text{(XIV.3)}$$

in which the current operator $J_a(\mathbf{q},i\omega_n)$ is written as

$$J_a(\mathbf{q},i\omega_n)=\frac{1}{\beta L^3}\sum_{ik_n}\sum_{\mathbf{k}\sigma}\frac{(2\mathbf{k}+\mathbf{q})_a}{2m}c^+_{\mathbf{k},ik_n\sigma}c_{\mathbf{k}+\mathbf{q},ik_n+i\omega_n\sigma} \quad \text{(XIV.4)}$$

Diagrammatically, defining Fermionic 4-momentum $\mathrm{k}\equiv(\mathbf{k},ik_n)$ , and Bosonic 4-momentum $\mathrm{q}\equiv(\mathbf{q},i\omega_n)$ , and further defining the bare two-terminal vertex function $\Gamma_0(\mathrm{k},\mathrm{k}+\mathrm{q})=(2\mathbf{k}+\mathbf{q})_a/2m$ and the Full Green's function $G(\mathrm{k})$ as

$\Gamma_{0,a}(\mathrm{k},\mathrm{k}+\mathrm{q})=$ 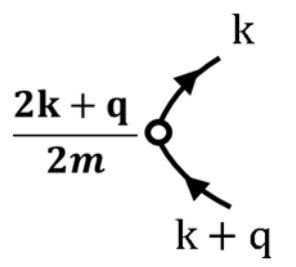

$G(\mathrm{k})=$ 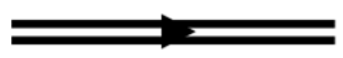

the current-current correlation function tensor can then be written as [83]

$\Pi_{ab}(\mathrm{q})=$ 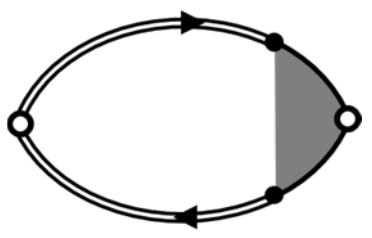

Written in equation form, we have

$$\Pi_{ab}(\mathrm{q})=-\frac{1}{\beta L^3}\sum_{\mathrm{k}}\begin{matrix}\Gamma_{0,a}(\mathrm{k},\mathrm{k}+\mathrm{q})G(\mathrm{k}+\mathrm{q})\\ \times G(\mathrm{k})\Gamma_b(\mathrm{k}+\mathrm{q},\mathrm{k})\end{matrix} \quad \text{(XIV.5)}$$

where $\sum_{\mathrm{k}}=\sum_{\mathbf{k}}\sum_{ik_n}$ , and $\Gamma(\mathrm{k}+\mathrm{q},\mathrm{k})$ is the dressed current vertex function depending on the detailed interaction. When considering the dislocation array scattering, the dressed vertex function $\Gamma(\mathrm{k}+\mathrm{q},\mathrm{k})$ under full Born approximation can be written as

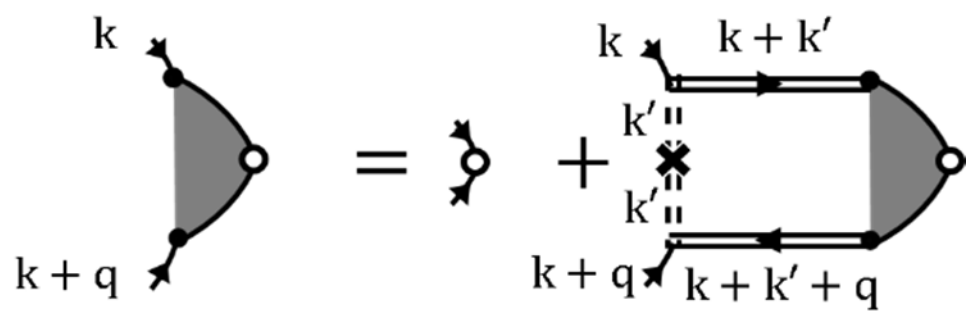

$$\Gamma_b(\mathrm{k}+\mathrm{q},\mathrm{k})=\Gamma_{0,b}(\mathrm{k}+\mathrm{q},\mathrm{k})+\frac{1}{\beta L^2}n_{dis}\times$$
$$\sum_{\mathbf{k}'}\left|\frac{A(\mathbf{k}')}{\varepsilon(\mathbf{k}')}\right|^2 G(\mathrm{k}+\mathrm{k}'+\mathrm{q})G(\mathrm{k}+\mathrm{k}')\Gamma_b(\mathrm{k}+\mathrm{k}'+\mathrm{q},\mathrm{k}+\mathrm{k}') \quad \text{(XIV.6)}$$

where $A(\mathrm{k}')=A(\mathbf{k}')\delta_{k'_n=0}\delta_{k_z=0}$ since it denotes the scattering strength with a quenched dislocation array, and $\varepsilon(\mathrm{k}')$ is the dielectric function screening the interaction. We should also keep in mind that the summation does not contain summation over fermionic Matsubara frequency, since there is no dynamic inelastic scattering in this process.

We now further compute the DC conductivity of electron – dislocation array scattering as a function of temperature. By taking the $\mathbf{q}\to 0$ limit, Eq. (XIV.5) can be further simplified as

$$\Pi_{ab}(\mathbf{q}\to 0;i\omega_n)=\frac{1}{L^3}\sum_{\mathbf{k}}\Gamma_{0,a}(\mathbf{k},\mathbf{k})S_b(\mathbf{k};i\omega_n) \quad \text{(XIV.7)}$$

where we have defined that

$$\begin{aligned}S_b(\mathbf{k};i\omega_n)&\equiv-\frac{1}{\beta}\sum_{k_n}f(ik_n,ik_n+i\omega_n)\\ &=-\frac{1}{\beta}\sum_{k_n}\begin{matrix}G(\mathbf{k};ik_n)G(\mathbf{k};ik_n+i\omega_n)\\ \times\Gamma_b(\mathbf{k},\mathbf{k};ik_n+i\omega_n,ik_n)\end{matrix}\end{aligned} \quad \text{(XIV.8)}$$

Now, using the residual theorem and Matsubara frequency summation technique, we have

$$S(\mathbf{k};i\omega_n)=\int_c\frac{dz}{2\pi i}n_F(z)f(z,z+i\omega_n) \quad \text{(XIV.9)}$$

since the Fermionic Matsubara frequencies $ik_n$ are the poles of the Fermi-Dirac distribution function $n_F(z)$ . Further computing the contour integral, noticing the branch cuts caused by two Green's functions, we have

$$\begin{aligned}&S(\mathbf{k};i\omega_n)=\\ &\int_{-\infty}^{+\infty}\frac{d\varepsilon}{2\pi i}n_F(\varepsilon-i\omega_n)\times\begin{bmatrix}f(\varepsilon-i\omega_n,\varepsilon+i\delta)-\\ f(\varepsilon-i\omega_n,\varepsilon-i\delta)\end{bmatrix}\\ &+\int_{-\infty}^{+\infty}\frac{d\varepsilon}{2\pi i}n_F(\varepsilon)\times\begin{bmatrix}f(\varepsilon+i\delta,\varepsilon+i\omega_n)-\\ f(\varepsilon-i\delta,\varepsilon+i\omega_n)\end{bmatrix}\end{aligned} \quad \text{(XIV.10)}$$

Now performing analytical continuation to real frequency $i\omega_n\to\omega+i\delta$ , and noticing that $n_F(\varepsilon-i\omega_n)=n_F(\varepsilon)$ , we have

$$\begin{aligned}&S^R(\mathbf{k};\omega)\\ &=\int_{-\infty}^{+\infty}\frac{d\varepsilon}{2\pi i}\begin{bmatrix}n_F(\varepsilon+\omega)-\\ n_F(\varepsilon)\end{bmatrix}f(\varepsilon-i\delta,\varepsilon+\omega+i\delta)\\ &+\int_{-\infty}^{+\infty}\frac{d\varepsilon}{2\pi i}\begin{bmatrix}n_F(\varepsilon)f(\varepsilon+i\delta,\varepsilon+\omega+i\delta)-\\ n_F(\varepsilon+\omega)f(\varepsilon-i\delta,\varepsilon+\omega-i\delta)\end{bmatrix}\end{aligned}$$

Noticing that $f^*(a+bi)=f(a-bi)$ and $n_F(\varepsilon+\omega)-n_F(\varepsilon)$ $\approx\omega\frac{\partial n_F(\varepsilon)}{\partial\varepsilon}$ , we have

$$\operatorname{Im} S^R(\mathbf{k};\omega) = \omega \operatorname{Re}\int_{-\infty}^{+\infty} \frac{d\varepsilon}{2\pi}\frac{\partial n_F(\varepsilon)}{\partial \varepsilon} \times \begin{bmatrix} f(\varepsilon - i\delta, \varepsilon+\omega+i\delta) - \\ f(\varepsilon + i\delta, \varepsilon+\omega+i\delta) \end{bmatrix} \quad \text{(XIV.11)}$$

Finally, by substituting Eq. (XIV.11) and (XIV.7) back to Eq. (XIV.1), the DC electrical conductivity of electron – dislocation array scattering can be written as

$$\operatorname{Re}\sigma_{ab}(\mathbf{q}\to 0, \omega \to 0) = e^2 \frac{1}{L^3}\sum_{\mathbf{k}} \Gamma_{0,a}(\mathbf{k},\mathbf{k}) \operatorname{Re}\int_{-\infty}^{+\infty} \frac{d\varepsilon}{2\pi}\frac{\partial n_F(\varepsilon)}{\partial \varepsilon} G(\mathbf{k};\varepsilon+i\delta) \times \begin{bmatrix} G(\mathbf{k};\varepsilon+i\delta)\Gamma_b(\mathbf{k},\mathbf{k};\varepsilon+i\delta,\varepsilon+i\delta) - \\ G(\mathbf{k};\varepsilon-i\delta)\Gamma_b(\mathbf{k},\mathbf{k};\varepsilon+i\delta,\varepsilon-i\delta) \end{bmatrix} \quad \text{(XIV.12)}$$

Now what remains is to compute the current vertex function $\Gamma$ and electron Green's function $G$. For vertex function, analytical continuing Eq. (XIV.6) back to real frequency, and shift variable $\mathbf{k}+\mathbf{k}' \to \mathbf{k}'$, we have the following self-consistent equation of the vertex function as

$$\Gamma_b(\mathbf{k},\mathbf{k};\varepsilon+i\delta,\varepsilon\pm i\delta) = \Gamma_{0,b}(\mathbf{k},\mathbf{k}) + \frac{n_{dis}}{L^2}\sum_{\mathbf{k}'}\left|\frac{A(\mathbf{k}'-\mathbf{k})}{\varepsilon(\mathbf{k}'-\mathbf{k})}\right|^2 \delta_{kz,k'z} G(\mathbf{k}';\varepsilon+i\delta) \times G(\mathbf{k}';\varepsilon\pm i\delta)\Gamma_b(\mathbf{k}',\mathbf{k}';\varepsilon+i\delta,\varepsilon\pm i\delta) \quad \text{(XIV.13)}$$

in which we have the non-interacting current vertex as $\Gamma_{0,b}(\mathbf{k},\mathbf{k}) = \frac{\mathbf{k}_b}{m}$ ($b = 1,2,3$ are Cartesian coordinates). The Green's functions in Eq. (XIV.13) can be written as

$$G(\mathbf{k}';\varepsilon\pm i\delta) = \frac{1}{\varepsilon - (E_{\mathbf{k}'} - \mu) - \Sigma(\mathbf{k},\varepsilon\pm i\delta) \pm i\delta} \quad \text{(XIV.14)}$$

where the self-energy $\Sigma(\mathbf{k},\varepsilon\pm i\delta)$ under the full Born approximation can be written as

$\Sigma(\mathbf{k},\varepsilon\pm i\delta) =$

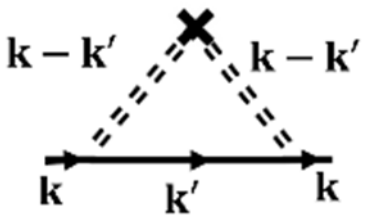

$$= \frac{n_{dis}}{L^2}\sum_{\mathbf{k}'} \frac{\left|A_{\mathbf{k}-\mathbf{k}'}/\varepsilon(\mathbf{k}-\mathbf{k}')\right|^2 \delta_{kz,k'z}}{\varepsilon - E_{\mathbf{k}'} + \mu \pm i\delta} \quad \text{(XIV.15)}$$

Now noticing the fact that $G(\mathbf{k};\varepsilon+i\delta)G(\mathbf{k};\varepsilon+i\delta) \ll G(\mathbf{k};\varepsilon+i\delta)G(\mathbf{k};\varepsilon-i\delta)$ [96], and using the identity that $G(\mathbf{k};\varepsilon+i\delta)G(\mathbf{k};\varepsilon-i\delta) \approx 2\pi\tau_{\mathbf{k}}\delta(\varepsilon - E_{\mathbf{k}} + \mu)$ for weak interaction, in which the relaxation time is defined as $\tau_{\mathbf{k}}^{-1} = -2\operatorname{Im}\Sigma(\mathbf{k},\varepsilon+i\delta)$, the DC conductivity of electron – dislocation array scattering Eq. (XIV.12) can be greatly simplified as

$$\operatorname{Re}\sigma_{ab}(T) = -\frac{e^2}{L^3}\sum_{\mathbf{k}} \Gamma_{0,a}(\mathbf{k},\mathbf{k}) \left.\frac{\partial n_F(\varepsilon)}{\partial \varepsilon}\right|_{E_{\mathbf{k}}-\mu} \tau_{\mathbf{k}} \times \operatorname{Re}\left[\Gamma_b(\mathbf{k},\mathbf{k};E_{\mathbf{k}}-\mu+i\delta,E_{\mathbf{k}}-\mu-i\delta)\right] \quad \text{(XIV.16)}$$

in which the relaxation rate under full Born approximation is given by

$$\frac{\hbar}{\tau_{\mathbf{k}}} = \frac{2\pi n_{dis}}{L^2}\sum_{\mathbf{k}'}\left|A_{\mathbf{k}-\mathbf{k}'}/\varepsilon(\mathbf{k}-\mathbf{k}')\right|^2 \delta_{kz,k'z}\delta(E_{\mathbf{k}} - E_{\mathbf{k}'}) \quad \text{(XIV.17)}$$

At this step, the temperature-dependent DC electrical conductivity tensor for electron – dislocation array scattering is complete: logically, to compute $\sigma_{ab}(T)$ using Eq. (XIV.16), one needs to first compute the current vertex function $\Gamma_b$, which is computed from Eq. (XIV.13). To compute Eq. (XIV.13), one further needs the electron Green's function, which is computed through Eqs. (XIV.14) and (XIV.15). The self-energy (XIV.15) contains the detailed information on electron-dislocation array interactions.

## XV. LATTICE THERMAL CONDUCTIVITY

It is well known that the dominant mechanism for phonon-dislocation scattering is the fluttering mechanism [29,32] instead of the phonon-dislon anharmonicity, which seemingly allows one to merely focus on the fluttering mechanism. However, for the real thermal conductivity calculation, the three-phonon anharmonic interaction still matters on top of the fluttering mechanism. Given the non-perturbative nature of the fluttering mechanism [36] (Eq. (XI.7)), it becomes impractical to adopt a perturbative approach to compute relaxation rate – the thermal conductivity should be computed directly using the phonon Green's function. In this section, we provide two equations to compute lattice thermal conductivity in a dislocated crystal: one is simpler, solely for fluttering mechanism, while the other is more involved but can take into account both the three-phonon and phonon-dislon anharmonic interactions together with the fluttering.

The generic thermal conductivity computed from the normal Green's function approach can be written as [97]

$$K(T) = \frac{k_B\beta}{3L^3}\lim_{\delta\to 0}\int_0^{+\infty} e^{-\delta t}dt\int_0^{\beta} d\lambda \left\langle \mathrm{S}(0)\cdot\mathrm{S}(t+i\lambda)\right\rangle \quad \text{(XV.1)}$$

where the energy flow vector operator $\mathrm{S}$ can be written as $\mathrm{S}(t) = \sum_{\mathbf{k}} \mathrm{v}_{\mathbf{k}}\omega_{\mathbf{k}} n_{\mathbf{k}}(t)$, where $n_{\mathbf{k}} = b_{\mathbf{k}}^{+}b_{\mathbf{k}}$ is the phonon number density operator, $\mathrm{v}_{\mathbf{k}}$ is the group velocity and $\omega_{\mathbf{k}}$ is the dispersion. Hence the central quantity to be computed is the 4-operator correlation $\left\langle b_{\mathbf{k}}^{+}(0)b_{\mathbf{k}}(0)b_{\mathbf{q}}^{+}(t+i\lambda)b_{\mathbf{q}}(t+i\lambda)\right\rangle$. Using Wick's theorem, since the fluttering mechanism (XI.7) is quadratic, the thermal average $\langle\ \rangle$ can be written as

$$\left\langle b_{\mathbf{k}}^{+}(0)b_{\mathbf{k}}(0)b_{\mathbf{q}}^{+}(t)b_{\mathbf{q}}(t)\right\rangle = \left\langle b_{\mathbf{k}}^{+}(0)b_{\mathbf{k}}(0)\right\rangle\left\langle b_{\mathbf{q}}^{+}(t)b_{\mathbf{q}}(t)\right\rangle + \left\langle b_{\mathbf{k}}^{+}(0)b_{\mathbf{q}}^{+}(t)\right\rangle\left\langle b_{\mathbf{k}}(0)b_{\mathbf{q}}(t)\right\rangle + \left\langle b_{\mathbf{k}}^{+}(0)b_{\mathbf{q}}(t)\right\rangle\left\langle b_{\mathbf{k}}(0)b_{\mathbf{q}}^{+}(t)\right\rangle \quad \text{(XV.2)}$$

Now, neglecting the equal-time correlation $\left\langle b_{\mathbf{k}}^{+}(0)b_{\mathbf{k}}(0)\right\rangle$ which does not contribute to transport, and neglecting $\left\langle b_{\mathbf{k}}^{+}(0)b_{\mathbf{q}}^{+}(t)\right\rangle$ which doesn't conserve particle number hence has negligible probability [97-99], only the 3rd term in Eq. (XV.2) is kept. The thermal conductivity in Eq. (XV.1) can be simplified as

$$K(T) = \frac{k_B\beta}{3L^3}\sum_{\mathbf{kq}} \mathrm{v}_{\mathbf{k}}\cdot\mathrm{v}_{\mathbf{q}}\omega_{\mathbf{k}}\omega_{\mathbf{q}} \lim_{\delta\to 0}\int_0^{+\infty} e^{-\delta t}dt \times \int_0^{\beta} d\lambda \left\langle b_{\mathbf{k}}^{+}(0)b_{\mathbf{q}}(t+i\lambda)\right\rangle\left\langle b_{\mathbf{k}}(0)b_{\mathbf{q}}^{+}(t+i\lambda)\right\rangle \quad \text{(XV.3)}$$

Now what remains is to compute the two-point correlation functions and link them to the fluttering action Eq. (XI.7). Here, we show that the key is to link through the spectral density function $A_{\mathbf{kq}}(\omega)$. First, similar to the electron case in Eq. (VIII.3), we define the Time-ordered phonon Green's function in imaginary time $D_{\mathbf{kq}}(\tau) = -\left\langle T_\tau[b_{\mathbf{k}}(\tau)b_{\mathbf{q}}^{+}(0)]\right\rangle$ and in the Matsubara frequency domain as

$$D_{\mathbf{kq}}(i\omega_l) = \int_0^{\beta} D_{\mathbf{kq}}(\tau)e^{+i\omega_l\tau}d\tau$$
$$D_{\mathbf{kq}}(\tau) = \frac{1}{\beta}\sum_n D_{\mathbf{kq}}(i\omega_l)e^{-i\omega_l\tau} \quad \text{(XV.4)}$$

where $\omega_l = 2\pi l k_B T$ ( $l = 0, \pm 1, \pm 2...$ ) are the Bosonic Matsubara frequencies. With this definition, the Matsubara frequency Green's function can be represented using Lehmann representation as

$$D_{\mathbf{kq}}(i\omega_l) = -\frac{1}{Z}\sum_{nm} \frac{\langle n|b_{\mathbf{k}}|m\rangle\langle m|b_{\mathbf{q}}^{+}|n\rangle}{E_n - E_m + i\omega_l} \times\left(e^{-\beta E_m} - e^{-\beta E_n}\right) \quad \text{(XV.5)}$$

where $|n\rangle$, $|m\rangle$ are the exact eigenstates of the full Hamiltonian with eigenvalues $E_n$ and $E_m$, respectively. The retarded and advanced Green's functions can be written as analytical continuation of the Matsubara frequency Green's function as $D_{\mathbf{kq}}^{R}(\omega) = D_{\mathbf{kq}}(i\omega_l \to \omega + i\delta)$

Now if we define the spectral density function $A_{\mathbf{kq}}(\omega)$

$$A_{\mathbf{kq}}(\omega) \equiv \frac{1}{Z}\sum_{mn} e^{-\beta E_m}\langle n|b_{\mathbf{k}}|m\rangle\langle m|b_{\mathbf{q}}^{+}|n\rangle \times\delta(\omega - E_m + E_n) \quad \text{(XV.6)}$$

we then have $\operatorname{Im} D_{\mathbf{kq}}^{R}(\omega) = -\pi(e^{\beta\omega}-1)A_{\mathbf{kq}}(\omega)$ valid, where we used the identity that $\operatorname{Im} 1/(x+i\delta) = -\pi\delta(x)$.

Using the same approach, we could also rewrite the correlation function $\left\langle b_{\mathbf{k}}^{+}(0)b_{\mathbf{q}}(t)\right\rangle$ (*t* in real time at this step) in terms of Lehmann representation as

$$\left\langle b_{\mathbf{k}}^{+}(0)b_{\mathbf{q}}(t)\right\rangle = \frac{1}{Z}\sum_{mn} e^{-\beta E_m}\langle n|b_{\mathbf{q}}|m\rangle \times \langle m|b_{\mathbf{k}}^{+}|n\rangle e^{-i(E_m-E_n)t} \quad \text{(XV.7)}$$

Then from Eqs. (XV.6) and (XV.7), we have

$$\left\langle b_{\mathbf{k}}^{+}(0)b_{\mathbf{q}}(t)\right\rangle = \int_{-\infty}^{+\infty} A_{\mathbf{qk}}(\omega)e^{-i\omega t}d\omega \quad \text{(XV.8)}$$

Similarly, we have

$$\left\langle b_{\mathbf{k}}(0)b_{\mathbf{q}}^{+}(t)\right\rangle = \frac{1}{Z}\sum_{nm} e^{-\beta E_n}\langle n|b_{\mathbf{k}}|m\rangle \times \langle m|b_{\mathbf{q}}^{+}|n\rangle e^{i(E_m-E_n)t} \quad \text{(XV.9)}$$

Then, using Eq. (XV.6) , we have

$$\left\langle b_{\mathbf{k}}(0)b_{\mathbf{q}}^{+}(t)\right\rangle = \int_{-\infty}^{+\infty} A_{\mathbf{kq}}(\omega)e^{+\beta\omega}e^{+i\omega t}d\omega \quad \text{(XV.10)}$$

Now substituting Eqs. (XV.8) and (XV.10) back to Eq. (XV.3), the thermal conductivity $K(T)$ is written as

$$K(T) = \frac{k_B\beta}{3L^3}\sum_{\mathbf{kq}} \mathrm{v}_{\mathbf{k}}\cdot\mathrm{v}_{\mathbf{q}}\omega_{\mathbf{k}}\omega_{\mathbf{q}} \lim_{\delta\to 0}\int_0^{+\infty} e^{-\delta t}dt\int_0^{\beta} d\lambda \times \int_{-\infty}^{+\infty} A_{\mathbf{qk}}(\omega')e^{-i\omega'(t+i\lambda)}d\omega'\int_{-\infty}^{+\infty} A_{\mathbf{kq}}(\omega)e^{+\beta\omega}e^{+i\omega(t+i\lambda)}d\omega \quad \text{(XV.11)}$$

Using the fact that $\delta(\omega'-\omega)\dfrac{e^{\beta(\omega-\omega')}-1}{\omega-\omega'} = \beta$, we have

$$K(T) = \frac{k_B\pi\beta^2}{3L^3}\sum_{\mathbf{kq}} \mathrm{v}_{\mathbf{k}}\cdot\mathrm{v}_{\mathbf{q}}\omega_{\mathbf{k}}\omega_{\mathbf{q}} \times\int_{-\infty}^{+\infty} d\omega e^{+\beta\omega}A_{\mathbf{qk}}(\omega)A_{\mathbf{kq}}(\omega) \quad \text{(XV.12)}$$

Finally, we obtain the thermal conductivity as

$$K(T) = \frac{k_B\beta^2}{3\pi L^3}\sum_{\mathbf{kq}} \mathrm{v}_{\mathbf{k}}\cdot\mathrm{v}_{\mathbf{q}}\omega_{\mathbf{k}}\omega_{\mathbf{q}} \times \int_{-\infty}^{+\infty} d\omega \frac{e^{+\beta\omega}}{(e^{\beta\omega}-1)^2}\operatorname{Im} D_{\mathbf{qk}}^{R}(\omega)\operatorname{Im} D_{\mathbf{kq}}^{R}(\omega) \quad \text{(XV.13)}$$

Eq. (XV.13) directly links the thermal conductivity $K(T)$ to the retarded phonon Green's function $D_{\mathbf{qk}}^{R}$, which can be obtained by analytical continuation of Matsubara Green's function as $D_{\mathbf{kq}}^{R}(\omega) = D_{\mathbf{kq}}(i\omega_l \to \omega + i\delta)$. As a simple check, assuming that the phonon Green's function can be written in a simplified form as

$$D_{\mathbf{qk}}^{R}(\omega) = \frac{\delta_{\mathbf{qk}}}{\omega - \omega_{\mathbf{k}} + i\gamma_{\mathbf{k}}(\omega)} \quad \text{(XV.14)}$$

where $\omega_{\mathbf{k}}$ is the phonon dispersion and $\gamma_{\mathbf{k}}(\omega)$ is the phonon spectral linewidth, we have

$$\operatorname{Im} D_{\mathbf{qk}}^{R}(\omega)\operatorname{Im} D_{\mathbf{kq}}^{R}(\omega) = \delta_{\mathbf{kq}}\left(\frac{\gamma_{\mathbf{k}}(\omega)}{(\omega-\omega_{\mathbf{k}})^2+\gamma_{\mathbf{k}}^2(\omega)}\right)^2 \approx \frac{\pi\delta_{\mathbf{kq}}\gamma_{\mathbf{k}}(\omega)\delta(\omega-\omega_{\mathbf{k}})}{(\omega-\omega_{\mathbf{k}})^2+\gamma_{\mathbf{k}}^2(\omega)}$$

The second equality is valid since the function is peaked at $\omega \approx \omega_{\mathbf{k}}$ and we have assumed an infinitesimal linewidth limit. Written in this way, the thermal conductivity Eq. (XV.13) can be further simplified as

$$\begin{aligned} K(T) &= \frac{k_B\beta^2}{3L^3}\sum_{\mathbf{k}} \mathrm{v}_{\mathbf{k}}^2\omega_{\mathbf{k}}^2 \frac{e^{+\beta\omega}}{(e^{\beta\omega}-1)^2}\frac{1}{\gamma_{\mathbf{k}}(\omega)} \\ &= \frac{1}{3}\sum_{\mathbf{k}} \mathrm{v}_{\mathbf{k}}^2\tau_{ph,\mathbf{k}}\mathrm{C}_{\mathbf{k}} \end{aligned} \tag{XV.15}$$

where $\tau_{ph,\mathbf{k}} = 1/\gamma_{\mathbf{k}}$ and $\mathrm{C}_{\mathbf{k}} = \omega_{\mathbf{k}}\partial/\partial T\left[n_B(\omega_{\mathbf{k}})\right]/L^3$ are phonon relaxation time and volumetric heat capacity, respectively, of a phonon with wavevector **k,** and $n_B(\omega_{\mathbf{k}})$ is the Bosonic occupation. Therefore, if the phonon relaxation rate $\gamma_{\mathbf{k}}(\omega)$ is known, Eq. (XV.13) can be rewritten in a much simpler but well-known form, where the 2nd line of Eq. (XV.15) is usually introduced in textbooks [100].

For fluttering mechanism action (IX.7), the Matsubara Green's function can be directly written as follows using Eq. (XI.7)

$$D_{\mathbf{kq}}(i\omega_n) = \frac{2D_{0n\mathbf{k}}\delta_{\mathbf{kq}}}{1 - D_{0n\mathbf{k}}J_{n\mathbf{k}} + \sqrt{1 + D_{0n\mathbf{k}}^2 J_{n\mathbf{k}}^2}} \tag{XV.16}$$

in which $D_{0n\mathbf{k}} \equiv \dfrac{1}{-i\omega_n + \hbar\omega_{\mathbf{k}}}$ is the free phonon propagator. This form goes beyond any perturbative expansion and cannot be written in terms of Eq. (XV.14), since the non-interacting phonon propagator $D_{0n\mathbf{k}}$ and interaction $J_{n\mathbf{k}}$ are highly entangled inside the " $\sqrt{\ }$ " symbol.

In real crystals, the contribution of thermal conductivity also contains three-phonon anharmonicity processes, where the coupling is the displacement field, which is proportional to $b_{\mathbf{k}} + b_{-\mathbf{k}}^+$. To facilitate the calculation with anharmonicity, we define another set of operators $X_{\mathbf{k}} \equiv b_{\mathbf{k}} + b_{-\mathbf{k}}^+ = X_{-\mathbf{k}}^+$ and $Y_{\mathbf{k}} \equiv b_{\mathbf{k}} - b_{-\mathbf{k}}^+ = -Y_{-\mathbf{k}}^+$, then we have

$$\begin{aligned} &\left\langle b_{\mathbf{k}}^+(t')b_{\mathbf{q}}(t)\right\rangle = \\ &\frac{1}{4}\begin{bmatrix} \left\langle X_{\mathbf{k}}^+(t')X_{\mathbf{q}}(t)\right\rangle + \left\langle X_{\mathbf{k}}^+(t')Y_{\mathbf{q}}(t)\right\rangle + \\ \left\langle Y_{\mathbf{k}}^+(t')X_{\mathbf{q}}(t)\right\rangle + \left\langle Y_{\mathbf{k}}^+(t')Y_{\mathbf{q}}(t)\right\rangle \end{bmatrix} \end{aligned} \tag{XV.17}$$

Then, if we define the correlation function for the displacement field in real time and frequency domain as

$$\begin{aligned} C_{\mathbf{qk}}(t-t') &= -i\theta(t-t')\left\langle\left[X_{\mathbf{q}}(t), X_{\mathbf{k}}^+(t')\right]\right\rangle \\ C_{\mathbf{qk}}(t-t') &= \int_{-\infty}^{\infty} C_{\mathbf{qk}}(\omega)e^{-i\omega(t-t')}d\omega \end{aligned} \tag{XV.18}$$

And following the same procedure of defining the spectral density function in Eqs. (XV.5)- (XV.10), we have the spectral function written as

$$\left\langle X_{\mathbf{k}}^+(t')X_{\mathbf{q}}(t)\right\rangle = \int_{-\infty}^{\infty}\Theta_{\mathbf{kq}}(\omega)e^{-i\omega(t-t')}d\omega \tag{XV.19}$$

where the spectral function also satisfies $\Theta_{\mathbf{kq}}(\omega) = -\dfrac{2}{\exp(\beta\hbar\omega)-1}\mathrm{Im}\,C_{\mathbf{qk}}^R(\omega)$, where $C_{\mathbf{qk}}^R(\omega)$ is the retarded correlation function $C_{\mathbf{qk}}^R(\omega) = C_{\mathbf{qk}}(\omega + i\delta)$. Now using the spectral function, following the same procedure of writing in terms of Lehmann representation, it can be proven that

$$\begin{aligned} \omega_{\mathbf{k}}\left\langle Y_{\mathbf{k}}^+(t')X_{\mathbf{q}}(t)\right\rangle &= \int_{-\infty}^{\infty}\omega\Theta_{\mathbf{kq}}(\omega)e^{-i\omega(t-t')}d\omega \\ \omega_{\mathbf{q}}\left\langle Y_{\mathbf{k}}^+(t')Y_{\mathbf{q}}(t)\right\rangle &= \frac{1}{\omega_k}\int_{-\infty}^{\infty}\omega^2\Theta_{\mathbf{kq}}(\omega)e^{-i\omega(t-t')}d\omega \\ \omega_{\mathbf{q}}\left\langle X_{\mathbf{k}}^+(t')Y_{\mathbf{q}}(t)\right\rangle &= \int_{-\infty}^{\infty}\omega\Theta_{\mathbf{kq}}(\omega)e^{-i\omega(t-t')}d\omega \end{aligned} \tag{XV.20}$$

Substituting Eqs. (XV.19) and (XV.20) back to Eq. (XV.17), we have

$$\begin{aligned} &\left\langle b_{\mathbf{k}}^+(0)b_{\mathbf{q}}(t)\right\rangle = \\ &\frac{1}{4}\int_{-\infty}^{\infty}\left(1 + \frac{\omega}{\omega_{\mathbf{k}}} + \frac{\omega}{\omega_{\mathbf{q}}} + \frac{\omega^2}{\omega_{\mathbf{k}}\omega_{\mathbf{q}}}\right)\Theta_{\mathbf{kq}}(\omega)e^{-i\omega t}d\omega \end{aligned} \tag{XV.21}$$

Similarly, we obtain

$$\begin{aligned} \left\langle b_{\mathbf{k}}(0)b_{\mathbf{q}}^+(t)\right\rangle = \frac{1}{4}\int_{-\infty}^{\infty} &\left(1 + \frac{\omega}{\omega_{\mathbf{k}}} + \frac{\omega}{\omega_{\mathbf{q}}} + \frac{\omega^2}{\omega_{\mathbf{k}}\omega_{\mathbf{q}}}\right) \\ &\times\Theta_{\mathbf{qk}}(\omega)e^{\beta\omega}e^{i\omega t}d\omega \end{aligned} \tag{XV.22}$$

Now if we substitute Eqs. (XV.21) and (XV.22) back to Eq. (XV.3), we finally have an alternative expression of thermal conductivity using the displacement-displacement correlation function as

$$\begin{aligned} K(T) = &\frac{\pi k_B\beta^2}{12L^3}\sum_{\mathbf{kq}}\mathrm{v}_{\mathbf{k}}\cdot\mathrm{v}_{\mathbf{q}}\omega_{\mathbf{k}}\omega_{\mathbf{q}}\int_{-\infty}^{\infty}d\omega\frac{e^{\beta\omega}}{\left(e^{\beta\omega}-1\right)^2} \\ &\times\left(1 + \frac{\omega}{\omega_q} + \frac{\omega}{\omega_k} + \frac{\omega^2}{\omega_k\omega_q}\right)^2\mathrm{Im}\,C_{\mathbf{qk}}^R(\omega)\,\mathrm{Im}\,C_{\mathbf{kq}}^R(\omega) \end{aligned} \tag{XV.23}$$

With Eqs. (XV.13) and (XV.23) in hand, we conclude this section.

## XVI. PERSPECTIVE

In this section, we briefly list a few possible research topics which can directly benefit from dislon theory. Due to our own limitations, this list is by no means exhaustive nor guaranteed to show significant effect. However, the dislon theory provides a tool to make such exploration possible. The readers are certainly encouraged to explore more opportunities using dislon theory for a better understanding of a dislocated crystal.

1) How do dislocations change materials' electronic structure, such as bandgaps in a semiconductor? This may be possible since recent study [101] has shown that large strain is capable of tuning the bandgap, while dislocations can introduce long-range strain fields.

2) Can dislocations induce Anderson localization? This may be possible since an analog of light localization using dislocations has been reported [102], not to mention the randomness that random dislocation lines can create.
3) Can dislocations drive a normal band insulator into a topological insulator? This may be realizable since dislocations can carry special topological states [103], and the tunable bandgap may result in an inversion between conduction and valence bands, which is the basis of forming a topological insulator [104].
4) Should dislocations increase or decrease a materials' heat capacity? Early simulation shows a very complicated scenario [105] which might be explained by dislon theory.
5) What is the role of dislocations in phase transition processes, including both first order transitions and continuous phase transitions? Early studies show the possible role of dislocations in the crystal melting process [106], while dislon theory may provide a comprehensive picture by considering other degrees of freedom in the crystal.
6) How can dislocations improve the thermoelectric figure-of-merit? Dislon theory may help quantify a few recent experimental findings, where dislocations have shown to dramatically increase the figure-of-merit in thermoelectric materials [107].
7) What is the interplay between dislocations and magnetic ordering? Recent experiment has shown a unique ferromagnetic ordering of a single dislocation in an antiferromagnetic material [63], which may be explained by a direct generalization of dislon theory with a spin degree of freedom – the observed ordering can be understood if a lower energy configuration is formed when a "spin dislon" starts to interact with the neighborhood magnetic environment.
8) If we do not restrict ourselves to dislocation-induced properties, since low-angle grain boundaries can be considered as an array of dislocations [108], the electronic properties of grain boundaries may be computed accordingly by constructing a dislon array. On the other hand, since infinitesimal dislocations have been applied to treat singularity of stress fields such as growth of fatigue cracks [109,110], the generalization of the present dislon theory may allow the calculation of electronic properties near a crack tip which has long been considered challenging.

There may also be opportunities to study the role of dislocations in nonlinear and quantum optics, chemical reactivity, and superconductivity. For instance, one recent application using the dislon theory shows that there are intrinsically two different types of electron-dislocation interactions, whose competition determines the critical temperature in a dislocated crystal [76].

One problem worth mentioning is an inverse problem to study how phonons and electrons affect dislocation motion. It is well known that the electrons in a superconductor can have a drag-like effect on dislocations, called electronic damping [111-114], while phonons can also influence the dislocation motion [79,115] which is examined in great detail in recent molecular dynamics simulations [116]. Since a dislocation's motion is slow compared to electronic and phononic processes, a pure classical and semi-classical theory is sufficient to describe dislocation motion without necessity to refer to a fully-quantized theory, although the mechanical properties can still be calculated from a Hamiltonian theory as done in density functional theoretical calculations [117].

## XVII. DISCUSSIONS AND CONCLUSIONS

In this study, we present a comprehensive theoretical framework of a quantized dislocation, namely a dislon, for arbitrary types of straight-line dislocations, and study its interplay with electronic and phononic degrees of freedom. Using this approach, one may be wondering how to check the validity of the dislon theory. In a few early studies of dislon theory, although the theoretical structure was simpler, we were able to perform a series of "sanity checks" to examine the validity of dislon theory:

1) Reducibility to classical displacement: The lattice displacement field can always be reduced to the well-known results normally introduced in classical materials science textbooks. For instance, this can be seen from Eq. (III.11).
2) The consistent deformation potential scattering. In Eq. (X.3), the electron-dislocation classical scattering amplitude $A_s$ , after a Fourier transform, gives $A(r) \propto \frac{b}{2\pi}\left(\frac{1-2\nu}{1-\nu}\right)\frac{\sin\theta}{r}$ , which is exactly the electron-dislocation deformation potential scattering form [118]. A detailed proof of this can be found in [76]. The benefit of the dislon approach is the elimination of any empirical parameters. In the classical deformation potential, there is an overall empirical prefactor in the expression of $A(r)$ [118], while the dislon theory fixes the prefactor with microscopic details.

3) Consistency of the electron-dislocation relaxation rate with, for instance, the semi-classical result [10]. The proofs can be found in [25,76].
4) The phonon-dislocation interaction beyond any perturbative approach. This non-perturbative nature can be seen directly from Eq. (XI.7), where the square root of the phonon Green's function prevents any perturbative analysis. In fact, this indicates the breakdown of *all* previously developed dislocation-phonon scattering theories, which are *all* based on perturbative theories. This conclusion, although seemingly drastic, has been confirmed by a recent and independent *ab initio* phonon-dislocation first-principles calculation of dislocation-phonon scattering, where the authors mentioned that "*Because of the breakdown of the Born approximation, earlier literature models fail, even qualitatively*"[80]. This independent first-principles calculation provided great confidence in the validity of dislon theory.
5) There are also a number of related features resulting from the non-perturbative nature of phonon-dislocation interaction. First, in normal perturbative studies, when a phonon is scatted by a dislocation, the dispersion of the phonon is considered a constant without change. The non-perturbative analysis shows that strong phonon-dislon interaction can change the phonon dispersion in an anisotropic way [36], which has been confirmed by independent *ab initio* calculations as well [80]. Moreover, perturbation theories showed that the dislocation-phonon relaxation rate varies monotonically with phonon frequency, whether for static scattering [26,27] or dynamic scattering [29,32], while the dislon theory predicts a resonance peak of relaxation rate for dislocation-phonon scattering. The existence of such resonance has also been confirmed by the same *ab initio* dislocation- phonon calculation [80].
6) The prediction of superconducting transition temperature $T_c$ in a dislocated superconductor. As seen in Eq. (X.3), there are two types of competing electron-dislocation interactions: classical and quantum. In one early study of dislon theory [76], we have shown that the $T_c$ is determined by this competition effect: when the quantum effect is dominant, then $T_c$ is increased, and vice versa. The computed $T_c$ shows good agreement with a number of existing experimental data; particularly, it may provide a feasible explanation of the mysterious dislocation-induced superconductivity [75,119].
7) A dislon is a quantized defect in crystalline solids. Recently, a quantized defect in liquid helium called "angulon" has been developed by Lemeshko *et al* [120-122]. Despite the fact that the dislon and angulon are completely different quasiparticles appearing in different environments, they both demonstrated the feasibility of actually quantizing a defect, which facilitates an understanding of the role of this very defect in a certain system – solid or liquid – using the many-body approach. For instance, in one recent angulon study [122], the phonon-angulon scattering can be readily studied.

Although the dislon theory has shown some early triumphs by explaining the non-perturbative phonon-dislocation interaction or superconductivity, it is still in an infant stage. For instance, in dislocated semiconductors [22,23] or ionic crystals [14], Coulomb scattering with charged dislocations becomes dominant due to the weak screening effect compared to metals. Such scattering has a pure classical nature and is related to the material-specific local atomic configuration near the dislocation core, but it has nothing to do with the dislocation's definition $\oint_D d\mathbf{u} = -\mathbf{b}$ , hence it hasn't been considered in this study. Even so, we feel that a direction generalization of Section III can still be applied to charged dislocations, magnetic dislocations, or disclinations, or even be used study more involved extended defects such as grain boundaries or nano-precipitates, which have both the "extended" part and "localized" part in space. Even so, we have no intention to treat an arbitrary dislocation such as a dislocation loop, where the expansion coefficient Eq. (III.11) becomes spatially dependent and an additional temporal variable is needed to track the position on the loop, which is the situation of string theory for a closed string and is mathematically formidable [123]. Moreover, as in dislon theory, where there are internal excitations as local vibrational modes, the internal excitation of other extended defects and their roles on materials' functional properties are worthwhile to explore. To facilitate future computation using this dislon theory, we provide a dimension analysis and a list of symbols (Appendices C and D). To conclude, we hope and believe that the present dislon theory could serve as a computational tool to help clarify the role of crystal dislocations on a number of functional properties of materials, at a new level of clarity.

**Acknowledgements**

The authors dedicate this paper to Prof. Mildred S. Dresselhaus who had greatly supported and supervised this research. M.L., Y.T. and G.C. thank support by S3TEC, an Energy Frontier Research Center funded by the U.S. Department of Energy (DOE), Office of Basic Energy Sciences (BES) under award No. DE-SC0001299/DE-FG02-09ER46577 (for fundamental research on electron phonon interaction in thermoelectric materials) and by the Defense Advanced Research Projects Agency (DARPA) MATRIX program HR0011-16-2-0041 (for developing and applying the simulation codes). Q.M. and Y.Z. thank the support by the U. S. DOE-BES Materials Science and Engineering Division, under Contract No. DE-AC02-98CH10886.

* Authors to whom correspondence should be addressed: mingda@mit.edu (M. L.); gchen2@mit.edu (G.C.)

## APPENDIX A. A FEW USEFUL IDENTITIES FOR SGN(K) FUNCTION

The vector sgn function is defined as

$$\mathrm{sgn}(\mathbf{k}) = \begin{cases} +1, & \text{if } k_x > 0 \\ -1, & \text{if } k_x < 0 \\ \mathrm{sgn}(k_y, \kappa), & \text{if } k_x = 0 \end{cases},$$

$$\mathrm{sgn}(k_y, \kappa) = \begin{cases} 1, & \text{if } k_y > 0 \\ -1, & \text{if } k_y < 0 \\ \mathrm{sgn}\,\kappa, & \text{if } k_y = 0 \end{cases},$$

$$\mathrm{sgn}(\kappa) = \begin{cases} 1, & \text{if } \kappa > 0 \\ -1, & \text{if } \kappa < 0 \\ 0, & \text{if } \kappa = 0 \end{cases}$$

And a relevant 2D version

$$\mathrm{sgn}(\mathbf{s}) = \begin{cases} +1, & \text{if } k_x > 0 \\ -1, & \text{if } k_x < 0 \\ \mathrm{sgn}(k_y), & \text{if } k_x = 0 \end{cases}, \quad \mathrm{sgn}(k_y) = \begin{cases} 1, & \text{if } k_y > 0 \\ -1, & \text{if } k_y < 0 \\ 0, & \text{if } k_y = 0 \end{cases}$$

This particular way of defining the vector sgn function is one of many choices which will prove handy later based on the following Lemmas .

**Lemma A1.**

$$\mathrm{sgn}(-\mathbf{k}) = -\mathrm{sgn}(\mathbf{k}), \text{ for } \forall \mathbf{k};\ \mathrm{sgn}(\mathbf{k}) = 0 \textit{ iff } \mathbf{k} = 0$$
$$\mathrm{sgn}(-\mathbf{s}) = -\mathrm{sgn}(\mathbf{s}), \text{ for } \forall \mathbf{s};\ \mathrm{sgn}(\mathbf{s}) = 0 \textit{ iff } \mathbf{s} = 0 .$$

where "iff" means "if and only if". The proof is straightforward.

**Lemma A2.**

$$\{\mathrm{sgn}\,\mathbf{s} \ge 0 | \kappa = 0\} \subset \{\mathrm{sgn}\,\mathbf{k} \ge 0\} .$$

Where "$\{\ \}$" is the notation for set. "$X|C$" means set X satisfies condition C.

**Proof:**

$$\{\mathrm{sgn}\,\mathbf{s} \ge 0 | \kappa = 0\} = \begin{Bmatrix} k_x > 0,\ \forall k_y,\ \kappa = 0 \\ k_x = 0,\ k_y > 0,\ \kappa = 0 \\ k_x = 0,\ k_y = 0,\ \kappa = 0 \end{Bmatrix} \subset \begin{Bmatrix} k_x > 0,\ \forall k_y,\ \forall \kappa \\ k_x = 0,\ k_y > 0,\ \forall \kappa \\ k_x = 0,\ k_y = 0,\ \kappa \ge 0 \end{Bmatrix} .$$
$$= \{\mathrm{sgn}\,\mathbf{k} \ge 0\}$$

QED.

**Lemma A3.**

$$\{\mathrm{sgn}\,\mathbf{s} \ge 0 | \kappa = 0\} = \{\mathrm{sgn}\,\mathbf{k} \ge 0 | \kappa = 0\}$$

**Proof:** From Lemma B2,

$$\{\mathrm{sgn}\,\mathbf{s} \ge 0 | \kappa = 0\} = \begin{Bmatrix} k_x > 0,\ \forall k_y,\ \kappa = 0 \\ k_x = 0,\ k_y > 0,\ \kappa = 0 \\ k_x = 0,\ k_y = 0,\ \kappa = 0 \end{Bmatrix} = \{\mathrm{sgn}\,\mathbf{k} \ge 0 | \kappa = 0\}$$

.

QED.

## APPENDIX B. THE DERIVATION OF DYSON'S EQUATION

### B1. Heisenberg equation of motion of operator (EoM)

For an arbitrary operator in imaginary time $O(\tau)$, we have $\partial_\tau O(\tau) = \partial_\tau \left[ e^{\tau H} O e^{-\tau H} \right] = [H, O](\tau)$

### B2. EoM of Green's function in imaginary time

For Green's function $G(\mathbf{k}\tau; \mathbf{k}'\tau') \equiv -\left\langle T_\tau c_{\mathbf{k}}(\tau) c^+_{\mathbf{k}'}(\tau') \right\rangle$, we have

$$\begin{aligned} &\partial_\tau G(\mathbf{k}\tau; \mathbf{k}'\tau') = \\ &-\partial_\tau \left\langle \theta(\tau - \tau') c_{\mathbf{k}}(\tau) c^+_{\mathbf{k}'}(\tau') \right\rangle + \partial_\tau \left\langle \theta(\tau' - \tau) c^+_{\mathbf{k}'}(\tau') c_{\mathbf{k}}(\tau) \right\rangle \\ &= -\delta(\tau - \tau') c_{\mathbf{k}}(\tau) c^+_{\mathbf{k}'}(\tau) - \delta(\tau - \tau') c^+_{\mathbf{k}'}(\tau) c_{\mathbf{k}}(\tau) \\ &\quad -\left\langle T_\tau \left[ \partial_\tau c_{\mathbf{k}}(\tau) \right] c^+_{\mathbf{k}'}(\tau') \right\rangle \\ &= -\delta(\tau - \tau') \delta_{\mathbf{k}\mathbf{k}'} - \left\langle T_\tau [H, c_{\mathbf{k}}](\tau) c^+_{\mathbf{k}'}(\tau') \right\rangle \end{aligned}$$

Up to this step, the expression is generic.

Now if we assume that the Hamiltonian $H$ is quadratic, $H = \sum_{\mathbf{pq}} h_{\mathbf{pq}} c^+_{\mathbf{p}} c_{\mathbf{q}}$, which is the case of external potential scattering, including dislocation scattering, and use the identity: $[AB, C] = A\{B, C\} - \{A, C\}B$, then we have

$$\begin{aligned} [H, c_{\mathbf{k}}] &= \sum_{\mathbf{pq}} h_{\mathbf{pq}} [c^+_{\mathbf{p}} c_{\mathbf{q}}, c_{\mathbf{k}}] \\ &= \sum_{\mathbf{pq}} h_{\mathbf{pq}} \left( c^+_{\mathbf{p}} \{c_{\mathbf{q}}, c_{\mathbf{k}}\} - \{c^+_{\mathbf{p}}, c_{\mathbf{k}}\} c_{\mathbf{q}} \right) \\ &= -\sum_{\mathbf{pq}} h_{\mathbf{pq}} \delta_{\mathbf{pk}} c_{\mathbf{q}} = -\sum_{\mathbf{q}} h_{\mathbf{kq}} c_{\mathbf{q}} \end{aligned}$$

Then the EoM of Green's function can be written as

$$\partial_\tau G(\mathbf{k}\tau;\mathbf{k}'\tau') = -\delta(\tau-\tau')\delta_{\mathbf{kk}'} + \left\langle T_\tau \sum_{\mathbf{q}} h_{\mathbf{kq}} c_{\mathbf{q}}(\tau) c^+_{\mathbf{k}'}(\tau') \right\rangle$$
$$= -\delta(\tau-\tau')\delta_{\mathbf{kk}'} - \sum_{\mathbf{q}} h_{\mathbf{kq}} G(\mathbf{q}\tau;\mathbf{k}'\tau')$$

**B3. Green's function in Matsubara frequency domain**

Now we further define Green's function in the Matsubara frequency domain as

$$G(\mathbf{k},\mathbf{k}';p_n) = \int_0^\beta G(\mathbf{k}\tau;\mathbf{k}'\tau') e^{ip_n(\tau-\tau')} d\tau$$
$$G(\mathbf{k}\tau;\mathbf{k}'\tau') = \frac{1}{\beta}\sum_n G(\mathbf{k},\mathbf{k}';p_n) e^{-ip_n(\tau-\tau')}$$

Then the EoM of Green's function in B2 can be rewritten as

$$-ip_n G(\mathbf{k},\mathbf{k}';p_n) = -\delta_{\mathbf{kk}'} - \sum_{\mathbf{q}} h_{\mathbf{kq}} G(\mathbf{q},\mathbf{k}';p_n)$$

**B4. Dyson's equation**

Now, divide the Hamiltonian into a diagonal part $H_0$ with eigenvalues $\varepsilon_{\mathbf{k}}$, and off-diagonal part $H_I$ with matrix element $v_{\mathbf{kq}}$, and define Green's function associated with Hamiltonian $H_0$ as $G_0(\mathbf{k},\mathbf{k}';p_n)$; then, we have

$$-ip_n G_0(\mathbf{k},\mathbf{k}';p_n) = -\delta_{\mathbf{kk}'} - \varepsilon_{\mathbf{k}} G_0(\mathbf{k},\mathbf{k}';p_n) \Rightarrow$$
$$G_0(\mathbf{k},\mathbf{k}';p_n) = \frac{\delta_{\mathbf{kk}'}}{ip_n - \varepsilon_{\mathbf{k}}} \equiv G_0(\mathbf{k};p_n)$$

The EoM of Green's function in the Matsubara frequency can be written as

$$-ip_n G(\mathbf{k},\mathbf{k}';p_n) = -\delta_{\mathbf{kk}'} - \sum_{\mathbf{q}} \left(\varepsilon_{\mathbf{k}}\delta_{\mathbf{kq}} + v_{\mathbf{kq}}\right) G(\mathbf{q},\mathbf{k}';p_n)$$
$$\Rightarrow \left(ip_n - \varepsilon_{\mathbf{k}}\right) G(\mathbf{k},\mathbf{k}';p_n) = \delta_{\mathbf{kk}'} + \sum_{\mathbf{q}} v_{\mathbf{kq}} G(\mathbf{q},\mathbf{k}';p_n)$$
$$\Rightarrow G(\mathbf{k},\mathbf{k}';p_n) = G_0(\mathbf{k},\mathbf{k}';p_n) + \sum_{\mathbf{q}} G_0(\mathbf{k},\mathbf{q};p_n) v_{\mathbf{kq}} G(\mathbf{q},\mathbf{k}';p_n)$$
$$\Rightarrow G = G_0 + G_0 V G$$

which is the Dyson's equation.

## APPENDIX C. DIMENSION ANALYSIS

**Fundamental constants**

$[\hbar] = \mathrm{M}^{+1}\mathrm{L}^{+2}\mathrm{T}^{-1}$, $[e] = \mathrm{M}^{1/2}\mathrm{L}^{3/2}\mathrm{T}^{-1}$ (static Coulomb)

**External parameters**

$[\rho] = \mathrm{M}^{+1}\mathrm{L}^{-3}$, $[\lambda] = [\mu] = \mathrm{M}^{+1}\mathrm{L}^{-1}\mathrm{T}^{-2}$, $[\beta] = \mathrm{M}^{-1}\mathrm{L}^{-2}\mathrm{T}^{+2} = [H^{-1}]$

**Displacements**

$[\mathbf{u}(\mathbf{R})] = \mathrm{L}$, $[\mathbf{U}_{\mathbf{k}}] = \mathrm{L}^3$, $[\mathbf{u}_{ph}(\mathbf{R})] = \mathrm{L}$, $[\mathbf{u}_{ph,\mathbf{k}}] = \mathrm{L}^{5/2}$, $[u_{\mathbf{k}}] = 1$

$[F(\mathbf{k})] = [F(\mathbf{s})] = \mathrm{L}^3$

**Hamiltonians**

$[H] = [T] = [U] = \mathrm{M}^{+1}\mathrm{L}^2\mathrm{T}^{-2}$, $[T_{\mathbf{k}}] = \mathrm{M}^{+1}\mathrm{L}^3$, $[W_{\mathbf{k}}] = \mathrm{M}^{+1}\mathrm{L}^3\mathrm{T}^{-2}$

$[\Omega_{\mathbf{k}}] = \mathrm{T}^{-1}$, $[Z_{\mathbf{k}}/A_{\mathbf{k}}] \equiv 1$, $[m_{\mathbf{k}}] = \mathrm{M}^{+1}\mathrm{L}^2$

$[V_{\mathbf{q}}] = \mathrm{M}^{+1/2}\mathrm{L}^{+7/2}\mathrm{T}^{-1}$, $[eV_{\mathbf{q}}] = \mathrm{M}^{+1}\mathrm{L}^{+5}\mathrm{T}^{-2}$,

$[\rho_e(\mathrm{r})] = \mathrm{M}^{1/2}\mathrm{L}^{-3/2}\mathrm{T}^{-1}$, $[V_{ei}] = \mathrm{M}^{+1/2}\mathrm{L}^{+1/2}\mathrm{T}^{-1}$,

$[C_{\mathrm{s}}] = 1$, $[g_{\mathbf{k}}] = \mathrm{M}^{+1}\mathrm{L}^2\mathrm{T}^{-2} = [H]$

$[\boldsymbol{\varepsilon}_{\mathbf{k}}] = 1$, $[\mathbf{u}_{ph}(\mathbf{R})] = \mathrm{L}^1$, $[\mathbf{u}_{ph,\mathbf{k}}] = \mathrm{L}^{+5/2}$,

$[\mathbf{p}_{ph}(\mathbf{R})] = \mathrm{M}^{+1}\mathrm{L}^{-2}\mathrm{T}^{-1}$, $[\mathbf{p}_{ph,\mathbf{k}}] = \mathrm{M}^{+1}\mathrm{L}^{-1/2}\mathrm{T}^{-1}$

$[\frac{\partial^3 U}{\partial u^3}] = \mathrm{M}^{+1}\mathrm{L}^{-10}\mathrm{T}^{-2}$, $[V^{abc}_{\mathbf{k}_1\mathbf{k}_2\mathbf{k}_3}] = \mathrm{M}^{+1}\mathrm{L}^{-4}\mathrm{T}^{-2}$,

$[A(\mathbf{k}_1,\mathbf{k}_2,\mathbf{k}_3)] = \mathrm{M}^{+1}\mathrm{L}^{-1}\mathrm{T}^{-1}$

**Actions**

$[d_{\mathbf{k}n}] = [f_{\mathbf{k}n}] = [\psi_{\mathbf{p}n\sigma}] = b_{\mathbf{k}n} = \mathrm{M}^{-1/2}\mathrm{L}^{-1}\mathrm{T}^{+1} = [H^{-1/2}]$,

$[\rho_{n\mathbf{k}}] = \mathrm{M}^{-1}\mathrm{L}^{-2}\mathrm{T}^{+2} = [H^{-1}]$,

$[\Delta_{\mathbf{k}}] = \mathrm{M}^{+1}\mathrm{L}^{-1}\mathrm{T}^{-1}$

## APPENDIX D. LIST OF SYMBOLS IN ALPHABETICAL ORDER

$\mathrm{A_s}$ Classical electron-dislon scattering amplitude Eq.IX.3
A($\mathbf{k}_1$, $\mathbf{k}_2$, $\mathbf{k}_3$) Anharmonic coupling constant
$b_{\mathbf{k}}$ Phonon annihilation operator
$b_{\mathbf{k}}^+$ Phonon creation operator
$b, \bar{b}$ Phonon fields in coherent state form
**b** Burgers vector
$c_{ijkl}$ Stiffness tensor
$c_{\mathbf{k}\sigma}$ Electron annihilation operator
$c_{\mathbf{k}\sigma}^+$ Electron creation operator
$C_{\mathbf{k}}$ Lattice specific heat
$\mathrm{C_s}$ $\kappa$-independent constant for dislon's constraint
$\mathrm{C}_{\mathbf{qk}}$ Correlation function of phonon displacement field
$d_{\mathbf{k}}$ Annihilation operator of dislon *d*-field
$d_{\mathbf{k}}^+$ Creation operator of dislon *d*-field
$d$, $\bar{d}$ Dislon *d*-fields in coherent state form
*D* Arbitrary loop enclosing the dislocation line
*D*[] Functional measure
$D_{0n\mathbf{k}}$ Free phonon propagator
$D_{\mathbf{kq}}$ Time-ordered phonon propagator
*e* Electron charge
$E_{\mathbf{p}}$ Electron single-particle energy
$E_n$ Eigenenergy of a generic Hamiltonian
$f_{\mathbf{k}}$ Annihilation operator of dislon *f*-field
$f_{\mathbf{k}}^+$ Creation operator of dislon *f*-field
$f$, $\bar{f}$ Dislon *f*-field in coherent state form in Eq. XIII.8
$F_{\mathbf{k}}$ Generic fermionic operator
$\mathrm{g}_{\mathbf{k}}$ Electron-dislon coupling *constant*
*G Electron* Green's function
*H* Hamiltonian
$J_a$ *a*-th component of electrical current operator

$J_{n\mathbf{k}}$ Phonon-dislon fluttering coupling constant
$k_{TF}$ Thomas Fermi screening wavevector
**k** Crystal momentum
$K$ Thermal conductivity
$L$ Crystal size
$m$ Electron mass
$m_\mathbf{k}$ Momentum-dependent mass term in dislon Hamiltonian
$M_\mathbf{k}$ Coefficient appearing in Eq. III.2
$n_{dis}$ Dislocation number density
$n_\mathbf{k}$ Electron number operator
$n_{B(F)}$ Bose (Fermi-Dirac) distribution function
**n** Normal vector to slip plane
$N$ Number of total atoms in perfectly periodic crystal
$N_{\mathrm{dis}}$ Number of dislocations
$p_\mathbf{k}$ Canonical momentum of dislon
$p_n$ Fermionic Matsubara frequency
$\mathbf{p}_{ph}$ Canonical momentum of phonon
**r** Position vector in 2D Cartesian coordinate
$\mathbf{r}_0$ Position vector of the dislocation core
**R** Position vector in 3D Cartesian coordinate
$\mathbf{R}^0_j$ Atomic coordinate of $j$-th atom in a perfect crystal
$S$ Action
S Energy flow operator
$T$ Dislocation kinetic energy
$T_\mathbf{k}$ Coefficient of dislocation's kinetic energy
**u** Lattice displacement vector
$u_\mathbf{k}$ Scalar Fourier component of lattice displacement
$U$ Dislocation potential energy
$\mathrm{v}_\mathbf{k}$ Group velocity
$\mathrm{v}_{\mathrm{ph},\mathbf{k}}$ Phonon phase velocity
$V_{ei}$ Electron-ion Coulomb potential
$V_{eff}$ Effective electron-dislocation coupling constant
$V_\mathbf{k}$ Fourier component of electron-ion potential
$V^{abc}_{\boldsymbol{k}_1\boldsymbol{k}_2\boldsymbol{k}_3}$ Anharmonic coefficient in momentum space
$W_\mathbf{k}$ Coefficient of dislocation's potential energy
$X_\mathbf{k}$ Operator used to define phonon correlation function
$Y_\mathbf{k}$ Operator used to define phonon correlation function
$Z$ Partition function
$Z_\mathbf{k}$ Normalization constant for dislon quantization

*Greek letters*
$\beta$ Inverse temperature
$\gamma$ Linewidth of phonon propagator
$\Gamma_0$ Free electron current vertex function
$\Gamma_\mathrm{b}$ Full electron current vertex function, b$^{\mathrm{th}}$ component
$\Delta_\mathbf{k}$ anharmonicity coefficient
$\varepsilon$ Dielectric function
$\varepsilon_\mathbf{k}$ Electron single particle energy
$\boldsymbol{\varepsilon}_\mathbf{k}$ Polarization vector of phonon (which distinguish the electron energy by context)
$\Theta_{\mathbf{qk}}$ Phonon spectral density function
$\kappa$ $Z$-component of momentum
$\lambda$ Lamé 1$^{\mathrm{st}}$ parameter
$\Lambda_\mathbf{k}$ Phonon-dislon anharmonic coupling constant
$\mu$ Lamé 2$^{\mathrm{nd}}$ parameter
$\mu$ Chemical potential (same symbol with Lamé 2$^{\mathrm{nd}}$ parameter but distinguishable by context)
$\nu$ Poisson ratio
$\Xi_\mathbf{k}$ Dislon field amplitude coefficient prefactor
$\bar{\pi}$ $\pi$ Momentum component in phonon field
$\Pi^R$ Retarded current-current correlation function
$\rho$ Mass density
$\rho_e$ Charge density operator
$\sigma$ Electrical conductivity
$\Sigma$ Self-energy
$\tau$ Imaginary time
$\tau_\mathbf{k}$ Relaxation time
$\phi$, $\bar{\phi}$ Phonon displacement field in coherent state form
$\Phi$ Composite operator for phonon anharmonicity study
$\psi$, $\bar{\psi}$ Electron fields in coherent state form
$\omega_D$ Debye frequency
$\omega_\mathbf{k}$ Phonon dispersion relation
$\omega_n$ Bosonic Matsubara frequency
$\Omega_\mathbf{k}$ dislon dispersion relation
$\mathcal{L}$ Lagrangian